\documentclass[useAMS,usenatbib]{mn2e}
\usepackage{graphicx}
\usepackage[T1]{fontenc}
\usepackage{aecompl}

\title[Interacting supernovae and impostors: SN~2007sv]{Interacting supernovae and supernova impostors. \\ SN~2007sv: the major eruption of a massive star in UGC~5979}
\author[L.~Tartaglia et al.]{L.~Tartaglia$^{1,2}$\thanks{E-mail: leonardo.tartaglia@oapd.inaf.it}, A.~Pastorello$^{2}$, S.~Taubenberger$^{3}$, E.~Cappellaro$^{1}$, J.R.~Maund$^{4}$,
\newauthor
S.~Benetti$^{1}$, T.~Boles$^{5}$, F.~Bufano$^{6}$, G.~Duszanowicz$^{7}$, N.~Elias-Rosa$^{1,8}$, 
\newauthor
A.~Harutyunyan$^{9}$, L.~Hermansson$^{10}$, P.~H{\"o}flich$^{11}$, K.~Maguire$^{12}$, H.~Navasardyan$^{1}$, 
\newauthor
S.J.~Smartt$^{4}$, F.~Taddia$^{13}$, M.~Turatto$^{1}$. \\
$^{1}$INAF-Osservatorio Astronomico di Padova, Vicolo dell'Osservatorio 5, I-35122 Padova, Italy \\
$^{2}$Universit\'a degli Studi di Padova, Dipartimento di Fisica e Astronomia, Vicolo dell'Osservatorio 2, I-35122 Padova, Italy \\
$^{3}$Max-Planck-Institut f\"ur Astrophysik, Karl-Schwarzschild-Str. 1, D-85748 Garching, Germany \\
$^{4}$Astrophysics Research Centre, School of Mathematics and Physics, Queen's University Belfast, Belfast BT7 1NN, UK  \\
$^{5}$Coddenham Astronomical Observatory, Suffolk IP6 9QY, UK \\
$^{6}$Departamento Ciencias Fisicas, Universidad Andr\'es Bello, Santiago de Chile, Chile \\
$^{7}$Moonbase Observatory, Otto Bondes vag 43, SE-18462 Akersberga, Sweden \\
$^{8}$Institut de Ci\`encies de l'Espai (CSIC - IEEC), Facultat de Ci\`encies, Campus UAB, 08193 Bellaterra, Spain. \\
$^{9}$Fundaci\'on Galileo Galilei - INAF, Telescopio Nazionale Galileo, Rambla Jos\'eAnaFern\'andez P\'erez 7, E-38712 Bre\~na Baja, Tenerife, Spain \\ 
$^{10}$Sandvretens Observatory, Linn\'egatan 5A, SE-75332 Uppsala, Sweden  \\
$^{11}$Department of Physics, Florida State University, 315 Keen Building, Tallahassee, FL 32306-4350, USA \\
$^{12}$European Southern Observatory (ESO), Karl Schwarschild Strasse 2, D-85748, Garching bei M\"unchen, Germany \\
$^{13}$The Oskar Klein Centre, Department of Astronomy, Stockholm University, AlbaNova, 10691 Stockholm, Sweden }

\begin{document}

\pagerange{\pageref{firstpage}--\pageref{lastpage}} \pubyear{}

\label{firstpage} 

\maketitle

\begin{abstract}
We report the results of the photometric and spectroscopic monitoring campaign of the transient SN~2007sv. 
The observables are similar to those of type IIn supernovae, a well-known class of objects whose ejecta interact with pre-existing circum-stellar material. 
The spectra show a blue continuum at early phases and prominent Balmer lines in emission, however, the absolute magnitude at the discovery of SN~2007sv (M$_R = -$14.25 $\pm$ 0.38) indicate it to be most likely a supernova impostor. 
This classification is also supported by the lack of evidence in the spectra of very high velocity material as expected in supernova ejecta. 
In addition we find no unequivocal evidence of broad lines of $\alpha$- and/or Fe-peak elements. 
The comparison with the absolute light curves of other interacting objects (including type IIn supernovae) highlights the overall similarity with the prototypical impostor SN~1997bs. 
This supports our claim that SN~2007sv was not a genuine supernova, and was instead a supernova impostor, most likely similar to the major eruption of a luminous blue variable. 
\end{abstract}


\section{Introduction} \label{intro}
With the label of {\it supernova} (SN) {\it impostors} we refer to a class of objects showing luminous outbursts that mimic the behaviour of real supernovae \citep[SNe; see e.g.][]{2000PASP..112.1532V,2012ASSL..384..249V,2006MNRAS.369..390M,2009MNRAS.394...21D}. 
The most classical SN impostors are thought to be the eruptions of extragalactic luminous blue variables \citep[LBVs,][]{1994PASP..106.1025H}. 
They may experience eruptions with comparable energies as those of real SNe, but the stars survive the eruptive events. \\ 
LBVs are evolved, massive stars very close to the classical Eddington limit, showing irregular outbursts and, occasionally, even giant eruptions during which they lose massive portions of their H-rich envelope (up to a few solar masses per episode). 
In quiescence they are blue stars located in the so called `S Doradus Instability Strip' of the HR Diagram, namely in the luminosity-temperature range $-9 \le$ M$_{\rm{bol}}$ $\le -$11 and 14000~K $\le$ T$_{\rm{eff}}$ $\le$ 35000~K \citep{1989A&A...217...87W}. 
During eruptive episodes LBVs become redder and evolve with a roughly constant bolometric magnitude. 
However, it has been proposed that they may increase their bolometric luminosity during giant eruptions \citep{1994PASP..106.1025H}. 
Active LBVs show quite erratic variability and, sometimes, fast optical luminosity declines after the outburst, possibly because of prompt dust formation in the circum-stellar environment. 
Spectroscopic studies indicate that eruptions are accompanied with relatively high velocity winds, viz. a few hundreds km~s$^{-1}$. 
Their spectra share some similarity with those of type IIn SNe, with prominent and narrow hydrogen lines in emission \citep{2000PASP..112.1532V}. \\

SN impostors are believed to be extra-Galactic counterparts of the famous `Giant Eruption' of the galactic LBV $\eta$ Carinae in the mid-19th century. 
This, together with the eruption of P Cygni in the 17th century, are the only two major eruptions registered in the Milky Way in recent times. 
However, weaker eruptions were occasionally observed in the past either in the Milky Way (e.g. AG Car) or in the Local Group \citep[e.g. S~Doradus in the LMC,][]{1994PASP..106.1025H}. 
These nearby examples are fundamental to our understanding of the nature of eruptive phenomena since their physical parameters are well constrained, and give us the opportunity to demonstrate that these stars survive major eruptions. 

It has been argued that a connection may exist between interacting SNe and impostors, mainly based on the observed similarity in the spectra, although they usually have remarkably different photometric properties. Even more importantly, there is evidence (though debated) that LBVs or other massive stars may explode as real interacting SNe soon after major outbursts \citep[e.g.][]{2007Natur.447..829P,2013MNRAS.431.2599M,2013arXiv1306.0038M,2013Natur.494...65O,2013MNRAS.tmp.2960S} or, at least, that interacting SNe are connected with massive stars compatible with LBVs \citep[][and references therein]{kot06,2009Natur.458..865G}.

In this context we report the case of SN~2007sv. The transient was discovered on 2007 December 20.912 UT, and was located 6".9 West and 6".7 South of the centre of UGC~5979 \citep{2007CBET.1182....1D}. 
The detection was confirmed with an unfiltered CCD image by T.~Boles on 2007 December 25.971 UT, whilst there was no source detected at the position of SN~2007sv on an archive image taken on 2007 September 13.093 UT \citep{2007CBET.1182....1D}. \\

This article is organised as follows. 
In Section 2 we report comprehensive information about the host galaxy and in Sections \ref{photometry} and \ref{spectroscopy} we present the results of our photometric and spectroscopic observations, respectively. 
A discussion follows in Section \ref{discussion}, where we remark similarities and differences between 2007sv and other interacting events. 
Finally our main conclusions are summarized in Section \ref{conclusions}.

Hereafter we will refer to SN impostors reporting their names without the `SN' prefix, in order to emphasize their different nature compared to genuine SNe.  

\section{The Host Galaxy} \label{host}
The host galaxy, UGC~5979, is a low-contrast faint (with apparent {\it B} magnitude 15.93) dwarf galaxy without a visible nucleus. 
Dwarf galaxies are the most common galaxies in the universe. 
\cite{2001ApSSS.277..231G} considers all galaxies with abolute magnitude fainter than M$_{V} \simeq -$18 as dwarfs, while according to \cite{1994ESOC...49....3T} the limit usually is M$_{B}$ $\simeq -$ 16. 
According to their optical appearance they are classified into five different groups: dwarf irregulars (dIs), blue compact dwarfs (BCDs), dwarf ellipticals (dEs), dwarf spheroidals (dSphs) and dwarf spirals (dSs). 
However, this morphological classification is somewhat arbitrary, and the distinction between different classes is sometimes ambiguous\footnote{Further sub-classification is based on the revised de Vaucouleurs morphological classification introduced by \cite{1960ApJ...131..215V} and on the luminosity classification introduced by \cite{1960ApJ...131..215V} \citep[and extended by][]{1985sgcc.book.....C}).}. 

UGC~5979 is a diffuse (dI) galaxy\footnote{http://leda.univ-lyon1.fr/}, located at {\it RA} = 10:52:41.16 and {\it Dec} = +67:59:18.8~[J2000], with a radial velocity corrected for the Local Group infall onto the Virgo cluster of about 1376~km~s$^{-1}$ (z = 0.0045). 
From the above value of the recessional velocity, we infer a distance for UGC~5979 of about 18.85 $\pm$ 1.03~Mpc, resulting in an absolute magnitude of M$_B = -$15.5 (distance modulus $\mu$ $\simeq$ 31.38 $\pm$ 0.27~mag, adopting H$_0$ = 73~km~s$^{-1}$ Mpc$^{-1}$).\\ 

For the foreground Galactic extinction we assumed the value A$_V$=0.048 mag, as derived from the \cite{2011ApJ...737..103S} recalibration of the \cite{1998ApJ...500..525S} infrared-based dust maps available e.g. in NED\footnote{http://ned.ipac.caltech.edu}. 
We also adopt no additional host galaxy extinction contribution in the transient direction, since a detailed analysis of the spectra of 2007sv revealed no evidence of narrow absorptions of the NaID doublet at the recessional velocity of the host galaxy. \\

A rough estimate of the metallicity of the host galaxy can be obtained from the relation of \cite{2004A&A...425..849P}:
\begin{equation}
    12 + \log{\rm{(O/H)}} = 5.80(\pm0.017) - 0.139(\pm0.011)\rm{M}_B
\end{equation}
that links the integrated absolute {\it B}-band magnitude with the average oxygen abundance of the galaxy, providing a value of $\sim$ 8, which suggests that the environment may have a significantly sub-solar metallicity. 
A direct measurement of the host galaxy metallicity confirms this result.
We spectroscopically observed UGC~5979 with the Nordical Optical Telescope (NOT) equipped with ALFOSC$+$grism\#4. A 1.0" slit was placed on a bright H~II region at 19.5" (1.8~kpc) from SN 2007sv. After 1800 sec exposure, we obtained an optical spectrum with clear detection of narrow [O~III]~$\lambda$5007~\AA, [O~III]~$\lambda$4959~\AA, Balmer lines up to H$\gamma$, [N~II]~$\lambda$6584~\AA~and [S~II].
Via Balmer decrement we determined an extinction of $E(B-V)$~$=$~0.82~mag at the H~II region location and we corrected the spectrum accordingly. We measured the line fluxes by fitting them with Gaussians, as explained in detail in \cite{2013A&A...558A.143T}.
The detection of N~II, H$\alpha$, H$\beta$ and O~III lines allowed us to determine the oxygen abundance via strong line diagnostics, namely with the N2 and O3N2 methods \citep{2004MNRAS.348L..59P}.
Our results are log(O/H)$+$12 (N2)~$=$~8.01$-$8.1~dex and log(O/H)$+$12 (O3N2)~$=$~8.00$-$8.08~dex.
The quoted uncertainty is due to the error on the flux of N~II, which appears rather faint.
Sub-solar metallicities may be a rather common characteristic of SN impostor environments \citep[see][]{2014MNRAS.441.2230H} and we are currently investigating this issue for a large sample of impostor environments (Taddia et al. in prep).

\section{Photometry} \label{photometry}
\subsection{Data reduction and light curves}
Our photometric monitoring campaign started on December 30, 2007 and spanned a period of about 100 days. 
We also collected sparse observations (mostly unfiltered) from amateur astronomers. 
Information about the photometric data and the instruments used are reported in Table \ref{LightCurves}. \\

All data were pre-processed using standard procedures in \textsc{iraf}\footnote{\textsc{iraf} is distributed by the National Optical Astronomy Observatory, which is operated by the Associated Universities for Research in Astronomy, Inc., under cooperative agreement with the National Science Foundation.} including bias and flat field corrections. 
To measure the magnitudes we used a dedicated pipeline developed by one of us (E. C.), that consists of a collection of \textsc{python} scripts calling standard \textsc{iraf} tasks (through \textsc{pyraf}) and other specific analysis tools, in particular \textsc{sextractor} for source extraction and star/galaxy separation, \textsc{daophot} to measure the source magnitude via PSF fitting and \textsc{hotpants}\footnote{http://www.astro.washington.edu/users/becker/hotpants.html} for image difference with PSF match.
\begin{table*}
\begin{minipage}[adjusting]{175mm}
 \caption[Photometry of 2007sv]{Optical magnitudes of 2007sv and associated errors.}
 \label{LightCurves}
 \begin{tabular}{@{}cccccccccccccr@{}}
  \hline
      Date &       MJD &      U &   err   &      B &  err  &      V &   err &      R &   err &     I  &  err  & Instrument  \\
  \hline
  20070913 & 54356.09 &     -       &    -       &    -        &   -       &    -        &   -       & $>$18.8&     -    &    -       &   -      & SX MX7      \\
  20071220 & 54454.91 &     -       &    -       &    -        &   -       &    -        &   -       & 17.161 & 0.261 &    -       &   -      & SX MX7       \\
  20071227 & 54461.86 &     -       &    -       &    -        &   -       & 17.975 & 0.356 & 17.502 & 0.137 &    -        &   -      & MX916        \\
  20071230 & 54464.15 &     -       &    -       & 18.502 & 0.077 & 18.161 & 0.039 & 17.786 & 0.023 & 17.432 & 0.023 & AFOSC       \\
  20080104 & 54469.04 & 19.145 & 0.147   & 18.860 & 0.025 & 18.220 & 0.014 & 17.812 & 0.024 &    -       &    -      & RATCam     \\
  20080106 & 54471.08 & 19.211 & 0.057   & 18.914 & 0.042 & 18.271 & 0.017 & 17.877 & 0.018 & 17.350 & 0.020 & RATCam     \\
  20080108 & 54473.21 & 19.264 & 0.085   & 18.963 & 0.025 & 18.225 & 0.019 & 17.802 & 0.029 & 17.327 & 0.020 & RATCam     \\
  20080110 & 54475.14 &    -        &     -       & 19.044 & 0.066 & 18.219 & 0.027 & 17.863 & 0.028 & 17.329 & 0.026 & AFOSC       \\
  20080110 & 54475.20 & 19.348 & 0.076   & 19.064 & 0.024 & 18.252 & 0.015 & 17.776 & 0.030 & 17.338 & 0.017 & RATCam     \\
  20080112 & 54477.25 & 19.373 & 0.025   & 19.124 & 0.024 & 18.311 & 0.026 & 17.929 & 0.033 & 17.522 & 0.032 & ALFOSC     \\
  20080113 & 54478.00 & 19.420 & 0.081   & 19.166 & 0.027 & 18.319 & 0.016 & 17.832 & 0.032 & 17.382 & 0.035 & RATCam     \\
  20080116 & 54481.12 & 19.544 & 0.138   & 19.208 & 0.020 & 18.354 & 0.026 & 17.768 & 0.029 & 17.452 & 0.028 & RATCam     \\
  20080118 & 54483.20 & 19.933 & 0.213   & 19.392 & 0.102 & 18.359 & 0.034 & 17.951 & 0.127 &    -        &   -       & RATCam     \\
  20080120 & 54485.94 &     -      &     -       &     -       &    -      &    -        &    -     & 17.984 & 0.265 &    -        &   -       & SX MX7       \\
  20080128 & 54493.16 &     -      &     -       & 19.512 & 0.109 & 18.573 & 0.030 & 18.020 & 0.068 & 17.467 & 0.085 & CAFOS        \\
  20080128 & 54493.95 &     -      &     -       & 19.526 & 0.037 & 18.520 & 0.022 & 17.907 & 0.023 & 17.406 & 0.024 & RATCam      \\
  20080207 & 54503.93 &     -      &     -       & 19.885 & 0.036 & 18.696 & 0.021 & 18.030 & 0.025 & 17.514 & 0.032 & RATCam       \\
  20080209 & 54505.94 &     -      &     -       &     -       &   -       &    -       &    -      & 18.023 & 0.179 &     -       &   -       & Apogee Ap7  \\
  20080211 & 54507.85 &     -      &     -       &     -       &   -       &    -       &    -      & 18.178 & 0.210 &     -       &   -       & SBIG ST-7    \\
  20080213 & 54509.94 &     -      &     -       &     -       &   -       &    -       &    -      & 18.166 & 0.218 &     -       &   -       & SX MX7        \\
  20080301 & 54526.88 &     -      &     -       & 20.532 & 0.053 & 19.211 & 0.028 & 18.328 & 0.041 & 17.843 & 0.056 & RATCam       \\
  20080304 & 54529.01 &     -      &     -       &     -       &   -       &    -       &    -      & 18.599 & 0.262 &     -       &    -      & SX MX7        \\
  20080305 & 54530.98 &     -      &     -       & $>$20.4&  -       & 19.471 & 0.056 & 18.456 & 0.067 & 17.854 & 0.035 & CAFOS         \\
  20080331 & 54556.00 &     -      &     -       & $>$21.2&  -       & 20.803 & 0.057 & 19.455 & 0.064 & 18.936 & 0.049 & ALFOSC       \\
  20080401 & 54557.93 &     -      &     -       & $>$21.4&  -       & 21.074 & 0.047 & 19.745 & 0.044 & 19.177 & 0.184 & RATCam       \\
  \hline
 \end{tabular}

 \medskip
 The observations were carried out using the 2.56~m Nordic Optical Telescope (NOT) with ALFOSC and the 2~m Liverpool Telescope with RATCam (both located at the Roque de Los Muchachos, La Palma, Canary Islands, Spain), the Calar Alto 2.2~m Telescope with CAFOS (Sierra de Los Filabres, Spain) and the 1.82~m Copernico Telescope with AFOSC (Mount Ekar, Asiago, Italy). Additional observations (mostly unfiltered) were obtained by amateur astronomers. \\
 The magnitudes obtained from SX MX7, Apogee Ap7 and SBIG ST-7 were computed from unfiltered images, whose magnitudes were rescaled to R-band. \\
 MX916: 0.45~m f4.5 newtonian telescope with a MX916 CCD Camera, at Mandi Observatory (Pagnacco, Udine, Italy)       \\ 
 SX MX7: 0.32~m f/3.1 reflector and a Starlight Xpress MX716 CCD camera at Moonbase Observatory (Akersberga, Sweden) \\
 Apogee AP7: C-14 Celestron Schmidt Cassegrain reflector and an Apogee AP7 CCD camera (Suffolk, United Kingdom)      \\
 SBIG ST-7: 0.44~m f4.43 telescope with a SBIG ST-7 Dual CCD Camera at Sandvretens Observatorium (Uppsala, Sweden)   \\
\end{minipage}
\end{table*}

\begin{table*}
\begin{minipage}{180mm}
 \caption[Local standards]{Magnitudes of the stellar sequence used for the photometric calibration. The stars are shown in Figure \ref{findingchart}.}
 \label{localstandards}
 \begin{tabular}{@{}ccccccccccccc@{}}
 \hline
 label & ra~[J2000]   & dec~[J2000] & U      & err   & B      & err   & V      & err   & R      & err   & I      & err   \\
       & (hh:mm:ss)   & (dd:mm:ss)  &        &       &        &       &        &       &        &       &        &       \\
 \hline
 1     & 10:53:02.62 & 67:58:18.77 &    -   &   -   &    -   &    -  & 20.159 & 0.056 & 19.202 & 0.015 &    -   &   -   \\
 2     & 10:53:01.92 & 67:59:46.33 & 18.451 & 0.020 & 17.110 & 0.015 & 15.861 & 0.029 & 14.983 & 0.022 & 14.214 & 0.027 \\
 3     & 10:52:54.29 & 67:58:14.49 & 15.692 & 0.009 & 15.675 & 0.007 & 15.127 & 0.020 & 14.800 & 0.016 & 14.429 & 0.030 \\
 4     & 10:52:50.03 & 67:57:21.82 &    -   &   -   & 20.571 & 0.010 & 19.113 & 0.024 & 18.195 & 0.015 & 16.425 & 0.014 \\
 5     & 10:52:49.13 & 67:58:39.87 & 17.682 & 0.007 & 17.088 & 0.011 & 16.212 & 0.012 & 15.688 & 0.008 & 15.180 & 0.012 \\
 6     & 10:52:43.00 & 67:57:22.46 & 16.401 & 0.004 & 16.329 & 0.005 & 15.661 & 0.010 & 15.295 & 0.006 & 14.887 & 0.003 \\
 7     & 10:52:27.18 & 68:00:26.52 &    -   &   -   &    -   &   -   & 18.245 & 0.013 & 17.770 & 0.005 & 17.299 & 0.015 \\
 \hline
 \end{tabular}
\end{minipage}
\end{table*}

\begin{figure}
 \begin{center}
 \includegraphics[scale=0.32]{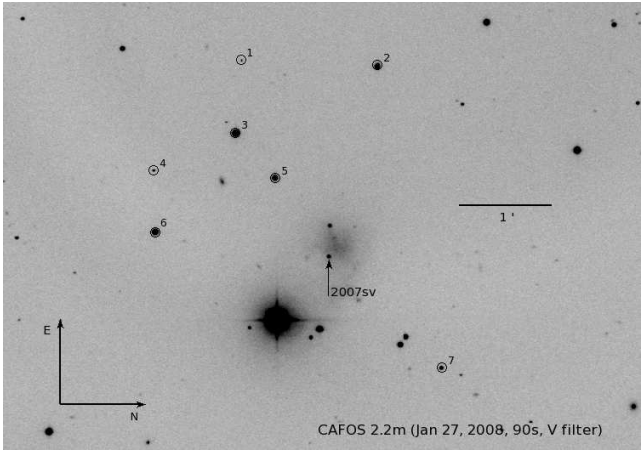}
 \caption[Findingchart for 2007sv]{Finding chart of 2007sv. The stars used for the photometric calibration are labelled with a number, the position of the transient is marked with an arrow. Information about the intrumental set-up is also reported.}
 \label{findingchart}
 \end{center}
\end{figure}

The magnitudes were measured using the PSF-fitting technique, first subtracting the sky background calculated using a low order polynomial fit (typically a 2nd order polynomial). 
The PSF was obtained by averaging the profiles of isolated field stars. 
The fitted source is removed from the original frames, then a new estimate of the local background is derived and the fitting procedure is iterated. 
Finally, the residuals are visually inspected to validate the fit. \\
Error estimates were obtained through artificial star experiments in which a fake star, of magnitude similar to that of the SN, is placed in the PSF-fit residual image in a position close to, but not coincident with that of the real source. 
The simulated image is processed through the PSF fitting procedure and the dispersion of measurements out of a number of experiments (with the fake star in slightly different positions), is taken as an estimate of the instrumental magnitude error. 
This is combined (in quadrature) with the PSF-fit error returned by \textsc{daophot}. \\
The SN photometry was calibrated as follows. Among the observational data, we selected the frames obtained in photometric nights in which standard photometric fields \cite[from the list of ][]{1992AJ....104..340L} were observed. 
These standard frames were used to derive zero points and colour terms for the specific instrumental set-up and to calibrate the magnitudes of selected stars in the SN field (Table \ref{localstandards} and Figure \ref{findingchart}). 
This local sequence was used to calibrate the SN magnitudes in non-photometric nights. 
The final magnitudes were computed using first-order colour-term corrections. 

\begin{figure}
 \begin{center}
 \includegraphics[scale=0.49]{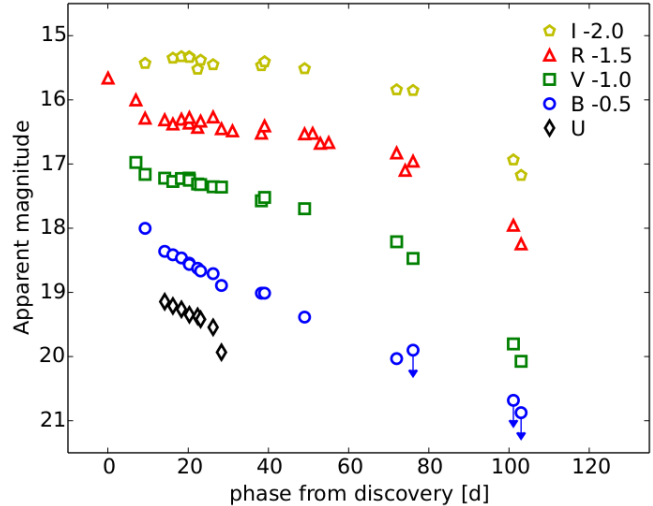}
 \end{center}
 \caption[lightCurve]{Multi-band light curves of 2007sv. The UBVRI magnitudes are listed in Table \ref{LightCurves}. The phases refer to the discovery.}
 \label{lightcurves}
\end{figure}

The resulting magnitudes of 2007sv are reported in Table \ref{LightCurves} along with the photometric errors, and the light curves are shown in Figure \ref{lightcurves}. 

\subsection{Constraining the SN age}
Since no observation was obtained a short time before the discovery of 2007sv, the epochs of the outburst on-set and the light curve maximum cannot be precisely constrained.
Nonetheless, we have adopted the epoch of the discovery as reference time for the light curve phases, assuming that  the transient was discovered in the proximity of the maximum light.
In other words, hereafter we will adopt as indicative epoch for the light curve maximum the discovery time (December 20, 2007; JD = 2454455.41).
The goodness of our assumption is supported by the photometric and spectroscopic analysis that will be widely discussed in the next sections.

Briefly, our choice is motivated on the basis of the following line of thinking. As we will discuss in the forthcoming sections, 2007sv shows very fast temperature/colour evolutions during the first
3-4 weeks, suggesting that the transient was discovered very young, a few days after the burst. This also implies that it is  unlikely that 2007sv  reached a peak magnitude much brighter than the  discovery magnitude $R$ = 17.16 (Table \ref{LightCurves}). If we  adopt for 2007sv the distance modulus and the reddening value discussed 
in Section \ref{host}, we  obtain an absolute magnitude  M$_{R} = -$14.25 $\pm$ 0.38 at discovery, which is an indicative estimate for the absolute peak magnitude. 
This weak absolute magnitude at maximum is an indication that 2007sv was very likely a SN impostor rather than a genuine SN explosion.

\subsection{Absolute magnitude and colour curves} \label{abscol}
\begin{figure}
 \begin{center}
 \includegraphics[scale=0.48]{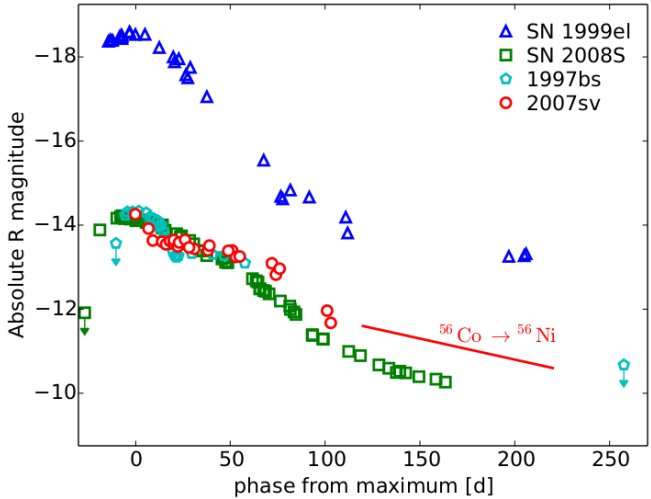}
 \end{center}
 \caption[Absolute R curves comparison]{Comparison of the absolute R-band light curves of the impostors 2007sv and 1997bs, the enigmatic transient SN~2008S, and the classical type IIn SN~1999el. The red line indicates the slope of the $^{56}$Co decay.}
 \label{abscurves}
\end{figure}

In Figure \ref{abscurves} we compare the absolute R-curve of 2007sv  with those of the SN impostor 1997bs \citep{2000PASP..112.1532V}, SN~2008S \citep{2009MNRAS.398.1041B} and the  type IIn SN~1999el \citep{2002ApJ...573..144D}. 
As highlighted in  this comparison, the  absolute peak magnitude of 2007sv is similar to those measured for the SN impostor 1997bs  and the enigmatic transient SN~2008S, and significantly fainter than that of a canonical SN IIn such as SN~1999el. 
Although the faint absolute magnitude supports the SN impostor scenario for 2007sv, this argument alone is not sufficient to rule out a true SN explosion, as we will discuss in Section 5. \\

\begin{figure}
 \begin{center}
 \includegraphics[scale=0.55]{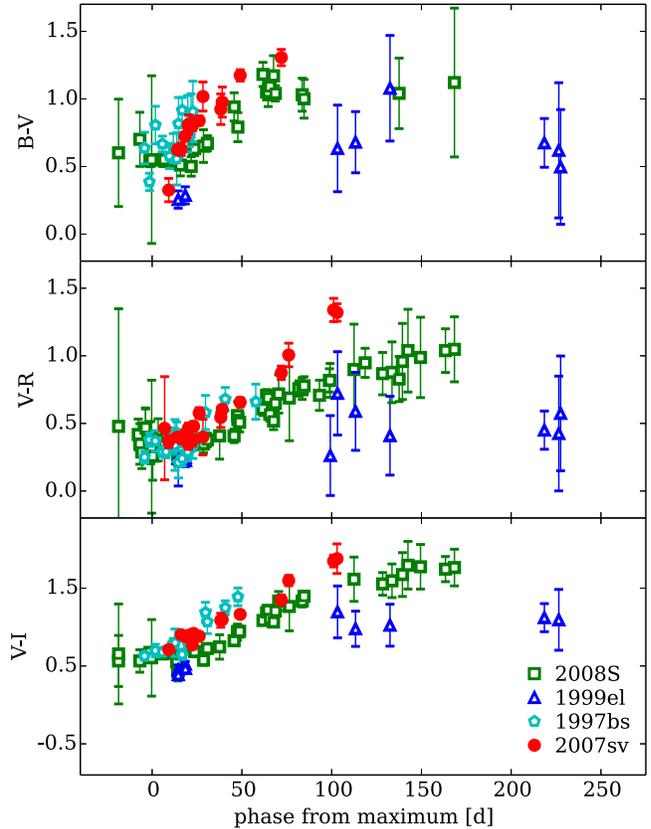}
 \caption[Colour-curves comparison]{Comparison among the $B-V$ (top), $V-R$ (middle) and $V-I$ (bottom) colour curves of the same sample as in Figure \ref{abscurves}. All phases refer to the epoch of the maximum, which for 2007sv we assumed to be coincident with the discovery epoch.}
 \label{colcurves}
 \end{center}
\end{figure}

The $B-V$, $V-R$ and $V-I$ colour curves of 2007sv are compared in Figure \ref{colcurves} with those of the same SNe considered in Figure \ref{abscurves}, showing that 2007sv rapidly becomes red (in analogy with other SN impostors discovered soon after the burst) as the temperature of the ejecta rapidly decreases (Section \ref{spectroscopy}). 
On the other hand, the regular type IIn SN~1999el remains bluer for a longer time. 
In particular, it has much bluer colours than the other transients at phases later than 100 days.

\begin{table*}
\begin{minipage}{175mm}
\caption[Log of the spectroscopical observations]{Log of the spectroscopical observations of 2007sv. The phase refers to the discovery.}
\label{speclog}
  \begin{tabular}{@{}cccccccc@{}}
  \hline
  Date & MJD & Phase  & Instrumental setup & Grism or grating & Spectral range  & Resolution  & Exp. times \\
       &     & (days) &                    &                  & (\AA)           & (\AA)       & (s)        \\
  \hline
  20071229 & 54463.96 & 9  & Ekar182+AFOSC  & 2$\times$gm4 & 3450-7800  & 24       & 2$\times$1800 \\
  20080105 & 54470.05 & 15 & TNG+LRS        & LR-B         & 3600-7770  & 19.9     & 2700          \\
  20080110 & 54475.08 & 20 & Ekar182+AFOSC  & gm2+gm4      & 3450-8100  & 34; 24   & 2$\times$1800 \\
  20080113 & 54478.26 & 23 & TNG+LRS        & LR-R         & 5070-10100 & 10.7     & 1800          \\
  20080114 & 54479.12 & 24 & NOT+ALFOSC     & gm4          & 3300-9100  & 9.6      & 1800          \\
  20080119 & 54484.43 & 30 & HET            & LRS          & 4300-7300  & 6.2      & 2$\times$1350 \\
  20080128 & 54493.17 & 38 & CAHA+CAFOS     & b200         & 3200-8800  & 12.3     & 3600          \\
  20080201 & 54497.12 & 42 & WHT+ISIS       & spec         & 3200-10300 & 4.8; 9.8 & 1200          \\
  20080301 & 54526.25 & 71 & HET            & LRS          & 4300-7300  & 6.0      & 4$\times$1125 \\
  20080308 & 54533.04 & 78 & CAHA+CAFOS     & b200+g200    & 4800-10700 & 13.8     & 2$\times$2400 \\ 
  \hline
  \end{tabular} 

\medskip
The spectra were obtained using the 1.82~m Telescopio Copernico with AFOSC, the 3.58~m Telescopio Nazionale Galileo (TNG) with DOLoRes (La Palma, Canary Islands, Spain), the 4.2~m William Herschel Telescope with ISIS (La Palma, Canary Islands, Spain), the 2.56~m Nordic Optical Telescope (NOT) with ALFOSC, the Calar Alto 2.2~m telescope with CAFOS and the 11.1x9.8~m Hobby-Eberly Telescope (HET, Mt. Fowlkes, Texas, USA) with LRS.

\end{minipage}
\end{table*}

The comparisons shown above highlight some of the similarities between the SN impostors 2007sv and 1997bs and the peculiar transient SN~2008S, whose nature has not been firmly established yet \citep[genuine electron-capture SN or SN impostor, see][and Section \ref{cfrsp}]{2008ApJ...681L...9P,2009MNRAS.398.1041B,2009ApJ...697L..49S,2009ApJ...705.1364T,2009ApJ...705L.138P,2010MNRAS.403..474W,2011ApJ...741...37K,2012ApJ...750...77S}, all showing a fainter maximum and a different evolution in the light curve when compared with the light curve of the interacting SN~1999el.

\section{Spectroscopy} \label{spectroscopy}
Spectroscopic observations were carried out from December 29, 2007 (i.e. 9 days after the discovery) to March 7, 2008 (+78 days from discovery). 
Basic information on the spectra and the instrumental configurations is reported in the log of spectroscopic observations (Table \ref{speclog}).

\begin{figure*}
\begin{center}
  \includegraphics[scale=0.95]{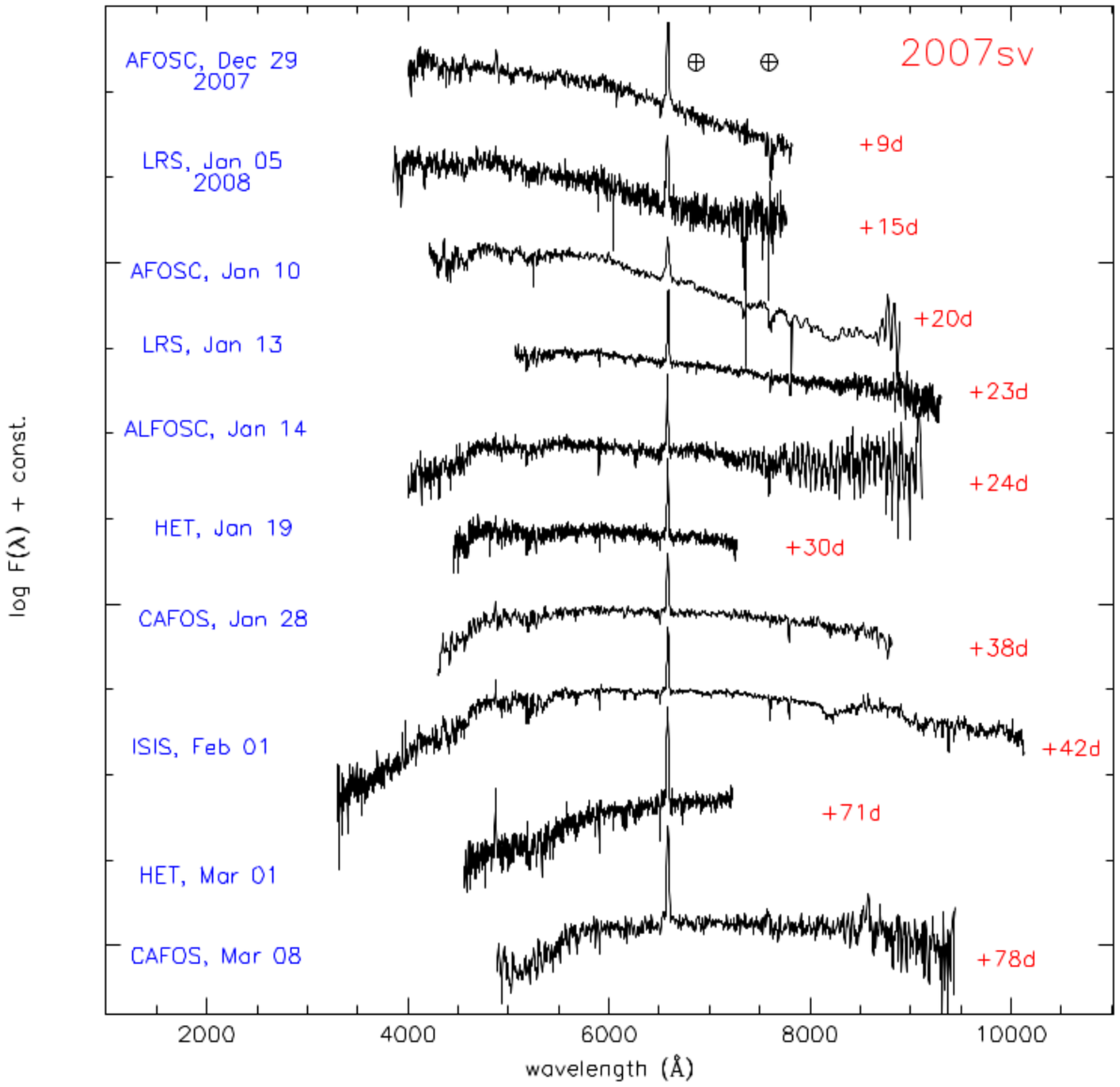}
  \caption[spectraSequence]{Spectral sequence of 2007sv. The date and the instrumental set-up are reported on the left, the phase (in days after the discovery) is indicated on the right. Spectra are flux-calibrated. The $\oplus$ symbols mark the position of the visible telluric absorptions.}
  \label{spectraSequence}
\end{center}
\end{figure*}

All data were processed using standard \textsc{iraf} tasks in order to perform the pre-reduction analysis (bias, flat and overscan corrections) and the extractions of the mono-dimensional spectra. 
Wavelength calibration was performed using the spectra of comparison lamps obtained with the same instrumental setup. 
Flux calibration was performed using the spectrum of standard stars. 
The accuracy of the wavelength calibration was verified measuring the wavelength of night sky lines (in particular [OI] at 5577.34~\AA~or 6300.30~\AA), and a shift was applied in case of discrepancy. 
Spectral resolution was measured from the full-width-at-half-maximum (FWHM) of night sky lines, adopting their mean value as final resolution estimate. 
The final spectral flux calibration was checked against multi-band photometry obtained on the nearest night and, when necessary, a scaling factor was applied. 
Telluric corrections were applied when spectra of references stars obtained during the same nights were available.
The spectral sequence obtained with the above procedures is shown in Figure \ref{spectraSequence}.

\begin{figure*}
\begin{center}
  \includegraphics[scale=0.57]{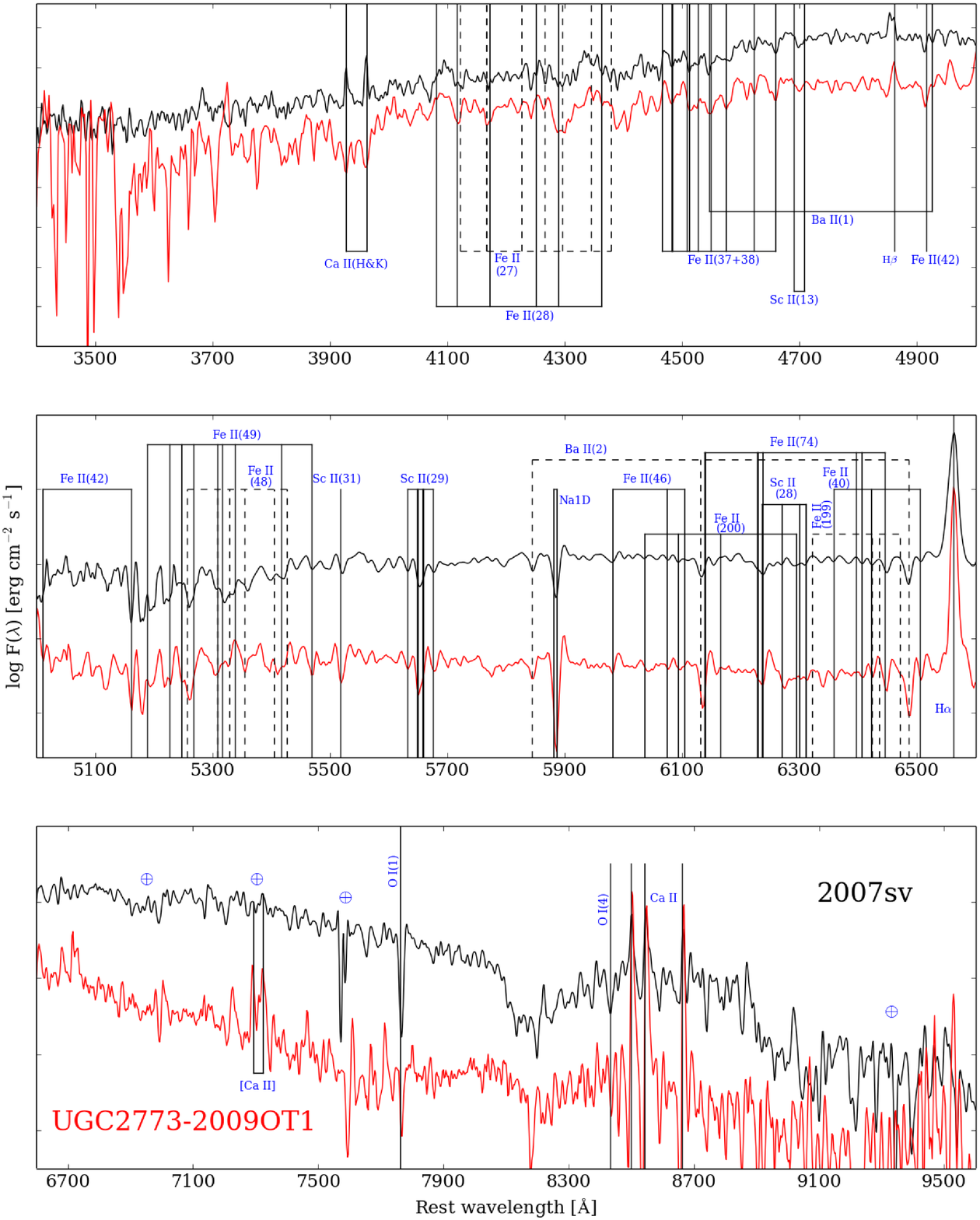}
  \caption[linesIdentification]{Line identification of the ISIS spectrum obtained on February 1, 2008 of 2007sv. A comparison with the spectrum of the transient UGC2773-2009OT1 obtained with the Telescopio Nazionale Galileo + LRS on Oct 11, 2009 (slightly before maximum, Padova-Asiago SN archive) is also shown. The spectra are flux-calibrated and redshift-corrected. H and Ca~II lines are marked at their rest wavelengths; the marks for the other lines are blue--shifted by $\simeq$500~km~s$^{-1}$. The $\oplus$ symbols mark the positions of prominent telluric absorption bands.}
  \label{lineid}
\end{center}
\end{figure*}

\subsection{Spectral evolution and line identification} \label{idlines}
From Figure \ref{spectraSequence}, we note that all spectra of 2007sv are dominated by a prominent and narrow H$_{\alpha}$ emission line. 
We also remark that there is relatively little evolution in the spectral features during the almost 80-days coverage window, except for the continuum becoming progressively redder. 
As the spectra are characterized by a significant continuum contribution to the total flux, we estimated the temperature of the emitting region through a black-body fit. 
The temperature experienced a rapid decline from $\simeq$ 8000~K in the +9d spectrum to $\simeq$ 5000 K at phase $\sim$20d.
Thereafter the temperature slowly declines to $\simeq$ 4000~K in the late-time spectra (71-78d; Figure \ref{spectraSequence}, see also Section \ref{hapro}).
This evolution is consistent with that of the broad-band colours (Figure \ref{colcurves}).
As mentioned above, there is little spectral evolution during the almost 80d of spectral coverage. However, it should be noticed that the spectra have relatively poor resolution and low S/N. Therefore, it is difficult to measure evolution in the weak and narrow spectral features.
In order to investigate the presence of low-contrast spectral lines, we inspected in detail one of our highest S/N spectrum (ISIS, phase +42d).

The line identification is shown in Figure \ref{lineid}.
The identification was performed by comparing the spectra of 2007sv with that of the SN impostor UGC2773-20009OT1 from the Padova-Asiago SN archive (also shown in Figures \ref{lineid} and \ref{speccompare}, res. 11~\AA). A comprehensive line identification for UGC2773-2009OT1 was performed by \citet{2010AJ....139.1451S} and \citet{2011ApJ...732...32F}.
Figure \ref{lineid} shows that, despite the different resolution, the spectra of the two transients are very similar, with a number of lines in common \citep[see e.g. Figures 8 to 12 in][]{2010AJ....139.1451S}. 
We determined an indicative photospheric velocity in the spectrum of 2007sv ($\simeq$~500~km~s$^{-1}$) from the blue-shifted absorption component of the Ba II 6496.6~\AA~line, and the expected positions of all other absorption lines were derived by adopting this velocity for all ions. 
The lines with strong emission components (e.g H$_{\alpha}$, H$_{\beta}$, H\&K Ca II and the NIR Ca II triplet) were identified using their rest wavelengths. 
We marked only multiplets with an intensity of the strongest line 5$\sigma$ above the noise level.
We also identified O~I (multiplets 1 and 4), Ba~II (multiplets 1 and 2), Na~ID (doublet at 5889.9 and 5895.9~\AA), Sc~II (multiplets 13, 28, 29, 31), Fe~II (multiplets 27, 28, 37, 38, 40, 42, 46, 48, 49, 74, 199, 200).
These lines show a narrow P-Cygni profile (blueshifted by about 650~km~s$^{-1}$), although a shallow high-velocity component in the NIR Ca II triplet cannot be ruled out (see Section \ref{cfrsp}).
We also marked the positions of the [Ca II] doublet lines at 7292\AA~and 7324\AA, which are prominent in the spectrum of UGC2773-20009OT1 but not in the 2007sv spectrum.\footnote{We also note that a telluric feature matches the wavelength position of the [Ca II] doublet, therefore no robust conclusion can be inferred on the identification of this feature in the 2007sv spectrum.}

\subsection{H$_\alpha$ profile and evolution of the main observables} \label{hapro}

\begin{figure}
\begin{center}
  \includegraphics[scale=0.44]{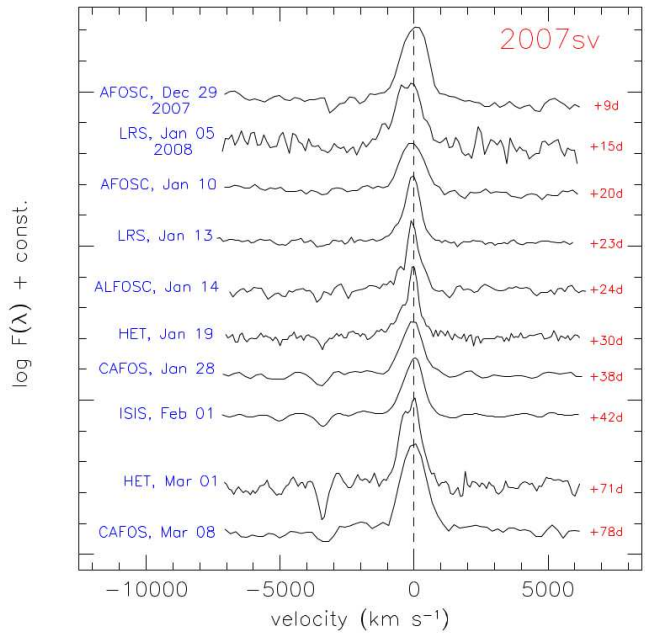}
  \caption[Evolution of the profile of H$_{\alpha}$]{Evolution of the profile of H$_{\alpha}$ in the velocity space.}
  \label{halphaev}
\end{center}
\end{figure}

The evolution of the H$_{\alpha}$ profile during the almost 80d of spectral coverage is highlighted in Figure \ref{halphaev}. 
We notice that there is no significant change in the wavelength position of the H$_{\alpha}$ emission peak.

The H$_{\alpha}$ line profile is relatively complex. 
A narrow component is detected in our higher resolution spectra of 2007sv, but a simple Gaussian or Lorentzian line fit does not well reproduce the entire line profile. 
For this reason, we adopted a combination of multiple line components to improve the accuracy of the spectral line fit. 
A broad component (decreasing from $\geq$ 2000 to $\simeq$ 1700~km~s$^{-1}$) is visible in the two earliest spectra (phases +9d and +15d). This component is only marginally detected in the two subsequent spectra (+20d and +23d), and disappears at later phases. In fact, the broad component is below the detection threshold in the +24d ALFOSC spectrum. 
Starting from the LRS spectrum at +23d, we improved our fits by including an intermediate-width (FWHM velocity $\approx$ 600). 
While we cannot rule out that the intermediate component was also present at earlier phases, the modest resolution of the +9d to +20d spectra prevents us its discrimination from the narrow component. 
In the spectra collected at later phases, a two-component (intermediate + narrow) fit well reproduces the observed line profile.

\begin{figure}
\begin{center}
  \includegraphics[scale=0.39]{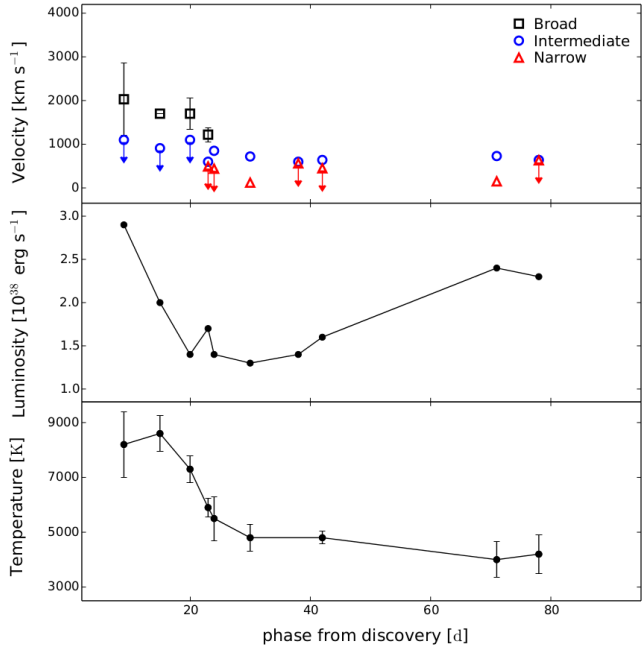}
  \caption[Velocity, $_\alpha$ luminosity and temperature evolution]{{\bf Top.} FWHM evolution for the broad (black squares, solid line), the intermediate (blue circles) and the narrow H$_{\alpha}$ (red triangles) components.
  {\bf Middle.} Evolution of the total luminosity of H$_{\alpha}$. We assume a 10$\%$ error in the measures, due only to the error in the flux calibration.
      {\bf Bottom.} Evolution of the spectral continuum temperature.}
  \label{tempspeed}
\end{center}
\end{figure}

In order to analyze the evolution of the velocity of the different H$\alpha$ components and the total line flux, we first corrected the spectra for redshift (adopting 1116~km~s$^{-1}$ as the mean heliocentric radial velocity\footnote{http://leda.univ-lyon1.fr/}) and for foreground Galactic extinction (using the values mentioned in Section \ref{host}). 
Then we measured the total line flux, and the full-width-at-half-maximum (FWHM) velocities of the three H$_{\alpha}$ line components. 
In most spectra the narrow component was unresolved, and even the intermediate component was occasionally below (or near) the resolution limits. 
In these cases, a multicomponent fit using a combination of Gaussian functions provided good fits. 
However, in some cases (namely for the two higher resolution HET spectra), we used a Gaussian function for the intermediate component and a Lorentzian profile for the narrow component. 
Figure \ref{tempspeed} (top panel) shows the evolution of the velocity of the ejected material for the three line components. 

As mentioned above, the narrow component was unresolved in most cases. 
When the narrow H$_{\alpha}$ was unresolved, we used the spectral resolution as an upper limit for the velocity of the slowest-moving material.  

When the narrow line component was resolved, we first corrected the measured FWHM for the spectral resolution ($width=\sqrt{\rm{FHWM}^2-\rm{res}^2}$) and then computed the velocity ($v=\frac{width}{6562.8} \times c$). 
In the highest resolution HET spectra at phases +30 and +71d we measured the FWHM of the narrower component as 120$\pm$30~km~s$^{-1}$ and 150$\pm$40~km~s$^{-1}$, respectively. 
The intermediate component remains at roughly constant velocity around 600-800~km~s$^{-1}$ at all epochs. 
Finally, the broad component is characterized by a fast decline from $\simeq$ 2000~km~s$^{-1}$ in our first spectrum to $\approx$ 1200~km~s$^{-1}$ in the +23d spectrum. 
We note that these values are significantly smaller that the typical values of $\simeq$ 10000~km~s$^{-1}$ measured in the ejecta of young SNe. \\

Multiple line components in the spectra of interacting objects are known to arise from different emitting gas shells \citep[see e.g.][]{1993MNRAS.262..128T}. 
The very small velocities inferred for the narrow H$_{\alpha}$ in the HET spectra (120-150~km~s$^{-1}$) are consistent with those expected in the winds of an LBV. 
The velocity evolution of the broad component is consistent with material violently ejected, and in particular with the velocities observed in the fastest hydrogen-rich material expelled in major eruptions of LBVs \citep{2008Natur.455..201S,2010MNRAS.408..181P,2013ApJ...767....1P}. 
More puzzling is the interpretation of the intermediate component. 
According to the interpretation usually adopted in interacting SNe \citep[see e.g.][]{1994ApJ...420..268C}, the intermediate velocity component arises in the gas region between the forward shock and the reverse shock. 
In the case of 2007sv, the relative strength of this component progressively increases with time with respect to that of the narrow component. 
This would support the idea that a significant fraction of the line flux at late phases arises from the gas interface between the two shock fronts, hence from the shocked gas region. 
In addition, one may note that the intermediate component is significantly blue-shifted with respects to the narrow one. 
In the ejecta/CSM interaction scenario, a blue-shifted intermediate component may be explained with an attenuation of the red line wind due to prompt dust formation in a post-shock cool dense shell, as observed in a number of interacting SNe \citep[e.g. 2006jc,][]{2008ApJ...680..568S,2008ApJ...684.1343N,2008MNRAS.389..141M}. 
Alternatively, very asymmetric and blue-shifted line profiles may be interpreted in terms of a highly asymmetric geometrical distribution of the CSM \citep[see e.g. the interpretation of][for SN~2006jd]{2012ApJ...756..173S}. \\

A progressive enhancement of ejecta/CSM interaction emission can be inferred observing the evolution of the total H$_{\alpha}$ flux in the latest spectra (phase $>$ 70d; Table \ref{specphysics} and Figure \ref{tempspeed}, middle panel). The flux decreases from about 7$\times$10$^{-15}$~erg~s$^{-1}$ to 3$\times$ 10$^{-15}$~erg~s$^{-1}$ during the first $\sim$ 40~days. 
Later on, we note an increase by a factor almost two in the H$_{\alpha}$ flux, approximatively about 5.5$\times$10$^{-15}$~erg~s$^{-1}$ in the last two spectra. As mentioned above, this can be interpreted as an increased contribution of the intermediate component arising in a shocked gas region which dominates the flux contribution at late phases over the other line components. 

\begin{table*}
\begin{minipage}{177mm}
 \caption[Main parameters inferred for the H$_{\alpha}$ line]{Main parameters inferred from the spectra of 2007sv.}
 \label{specphysics}
 \begin{tabular}{@{}cccccccc@{}}
 \hline
 Phase     & FWHM(H$_{\alpha,broad}$) & FWHM(H$_{\alpha,intermediate}$) & FWHM(H$_{\alpha,narrow}$) & H$_{\alpha}$ Luminosity & Resolution   \\ 
 (days)    & (km s$^{-1}$)         & (km s$^{-1}$)               & (km s$^{-1}$)          & (10$^{38}$ erg s$^{-1}$) & (km s$^{-1}$) \\  
 \hline
 9  & 2030$\pm$830 & $<$ 1100    & -          & 2.9 & 1100 \\
 15 & 1700:        & $<$ 910     & -          & 2.0 &  910 \\
 20 & 1700$\pm$360 & $<$ 1100    & -          & 1.4 & 1100 \\
 23 & 1220$\pm$160 & 600$\pm$100 & $<$ 490    & 1.7 &  490 \\ 
 24 & -            & 850$\pm$180 & $<$ 440    & 1.4 &  440 \\
 30 & -            & 720$\pm$100 & 120$\pm$30 & 1.3 &  280 \\
 38 & -            & 600$\pm$120 & $<$ 560    & 1.4 &  560 \\
 42 & -            & 640$\pm$100 & $<$ 450    & 1.6 &  450 \\
 71 & -            & 730$\pm$100 & 150$\pm$40 & 2.4 &  270 \\
 78 & -            & 640$\pm$130 & $<$ 630    & 2.3 &  630 \\
 \hline
 \end{tabular}

\medskip
The measured FWHM of the broad, intermediate and narrow components of H$_{\alpha}$ are reported in columns 2, 3 and 4, respectively. The measure marked with the : symbol is uncertain. \\ 
The total luminosity of H$_{\alpha}$ is in column 5, the spectral resolution in column 6.
\end{minipage}
\end{table*}

\subsection{Spectral comparison with other interacting transients} \label{cfrsp}

\begin{figure}
\begin{center}
  \includegraphics[scale=0.44]{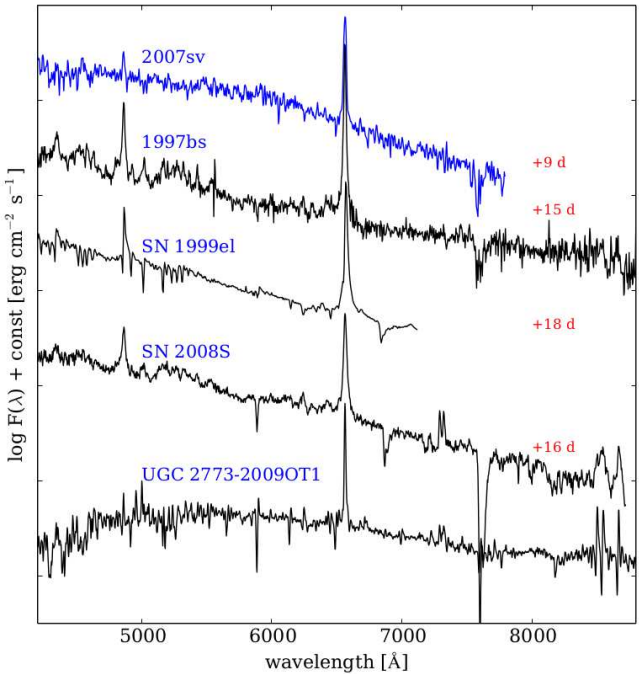}
  \caption[Spectral comparison]{Comparison of early-time spectra of the impostors 2007sv, 1997bs (unpublished spectrum from the Padova--Asiago SN archive obtained on 1997 April 30th with the ESO1.52~m telescope, resolution 10\AA, located at La Silla, Chile) and UGC2773-2009OT1 (spectrum obtained slightly before maximum), the enigmatic transient SN~2008S, and the linearly-declining type IIn SN~1999el. The spectra are flux-calibrated and redshift-corrected.}
  \label{speccompare}
\end{center}
\end{figure} 

An important issue is to determine whether the spectroscopy alone allows us to discriminate between genuine type IIn SNe and SN impostors. 
For this goal, we compare in Figure \ref{speccompare} the AFOSC early-time spectrum (phase +9d) of 2007sv with spectra of young transients with narrow emissions, viz. the impostors 1997bs \citep{2000PASP..112.1532V} and UGC2773-2009OT1 (Padova--Asiago SN Archive; Pastorello et al. in preparation), the classical type IIn SN~1999el \citep{2002ApJ...573..144D} and SN~2008S \citep{2009MNRAS.398.1041B}. 
SN~2008S is the prototype of a small family of intermediate-luminosity transients \citep[see][and references therein]{2009ApJ...705.1364T} whose nature has been widely debated. 
Although many observables of SN~2008S are similar to those observed in SN impostors, the detection of prominent, narrow [Ca~II] (7292--7324~\AA) lines and, even more, the late-time light curve with a decline rate consistent with that expected from the $^{56}$Co decay \citep{2009MNRAS.398.1041B}, provide reasonable arguments to support a faint SN scenario. 
The progenitor star of SN~2008S was detected in mid-infrared archive Spitzer images, whilst there was no detection in deep optical and near-IR pre-explosion frames \citep[e.g.][]{2008ApJ...681L...9P}. This was interpreted as a clear signature that the progenitor was a highly reddened star, embedded in a dusty environment. 
Although there is general agreement that the progenitor star of SN~2008S was a moderate-mass star\footnote{A similar conclusion was also inferred for the detected progenitor of the 2008S-analogous NGC300-2008OT1 \citep{2009ApJ...695L.154B,2009ApJ...699.1850B}}, the characterization of the stellar type is somewhat different in the different papers, ranging from a $\sim$9~M$_{\odot}$ extreme asymptotic giant branch star \citep[AGB; e.g.][]{2008ApJ...681L...9P,2010ApJ...715.1094K,2011ApJ...741...37K,2012ApJ...750...77S} to a $\leq$ 20~M$_\odot$ supergiant \citep[][]{2009ApJ...697L..49S}. 
The most debated issue is whether the observed 2008 outburst was a terminal stellar explosion, most likely as an electron-capture SN from a super-AGB star \citep{2009ApJ...705L.138P,2013ApJ...771L..12T} or an LBV-like outburst of a mildly massive star \citep[][]{2009ApJ...697L..49S,2011MNRAS.415..773S}. 

From the comparison in Figure \ref{speccompare}, it is evident that the spectra of all these transients are rather similar, and many spectral lines are in common to all of them. 
Therefore this is an indication that the spectra alone may not be sufficient to discriminate between impostors and true SNe. 
As mentioned before, the narrow [Ca~II] doublet at 7292--7324~\AA~is the hallmark feature for SN~2008S-like transients and is sometimes used as an argument to support the SN nature of these objects. 
We note that there is no clear evidence for the presence of [Ca~II] lines in the spectra of 2007sv or 1997bs. 
However, we have to admit that the [Ca~II] feature was detected in UGC2773-2009OT1 \citep[which is clearly an impostor, see][]{2010AJ....139.1451S,2011ApJ...732...32F}.
Therefore, the [Ca~II] feature is not a good discriminant of the nature of these explosions.

\begin{figure}
\begin{center}
  \includegraphics[scale=0.44]{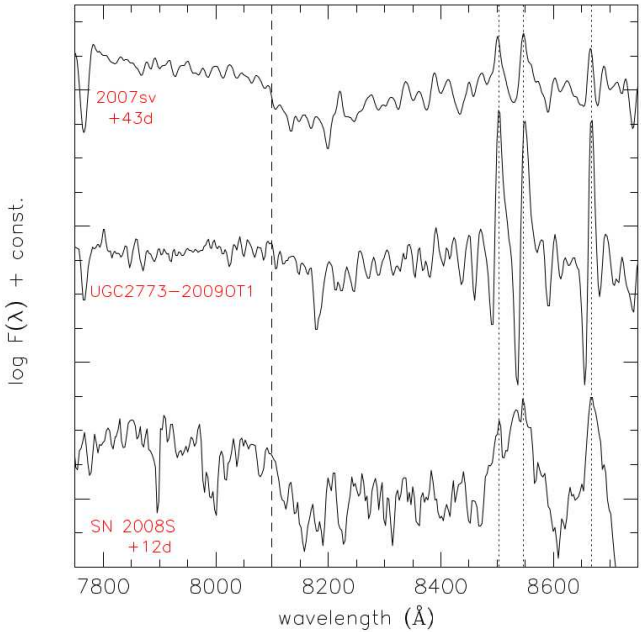}
  \caption[Comparison between Ca~II lines of different objects]{Comparison of the NIR Ca~II triplet line profiles in three different types of faint transients, viz. 2007sv, UGC2773-2009OT1 and SN~2008S (from top to bottom). The three dotted lines mark the position of the lines of the Ca~II triplet at 8498.0~\AA~8542.1~\AA~and 8662.1~\AA. The dashed is an indicative line that marks the velocity close to the terminal velocity of the gas. The spectra are flux-calibrated and redshift-corrected.}
  \label{cacomp}
\end{center}
\end{figure}

In Figure \ref{cacomp}, the 7800--8700~\AA~wavelength window of the +42d ISIS spectrum of 2007sv is compared with spectra of SN~2008S and UGC2773-2009OT1. 
In all of them we find the Ca~II triplet at 8498.0~\AA, 8542.1~\AA~and 8662.1~\AA, which is another very common feature in many types of transients, although some differences in the line strengths and velocities can be appreciated. 
The three spectra show narrow features with velocities of a few hundreds~km~s$^{-1}$, and these mark the presence of slow-moving material. 
However, a very broad depression with a minimum at about 8200~\AA~is visible in all spectra in Figure \ref{cacomp}, suggesting that a small amount of material can be ejected at high velocities (above 10000~km~s$^{-1}$) also in SN impostors. 
The presence of fast-moving material has been also reported in the $\eta$ Car circum-stellar environment \citep{2008Natur.455..201S}. 
Therefore, the detection of high-velocity gas alone should not be considered a robust argument to favor a SN scenario\footnote{We note, however, that this argument is instead used by many authors to favor the SN explosion scenario for the debated SN~2009ip.}.

\section{Discussion} \label{discussion}
In Sections 3 and 4 the photometric and spectroscopic properties of the optical transient 2007sv have been described. 
The main goal of the forthcoming discussion is to provide convincing insights on its nature (SN vs. SN impostor). 
Already from a quick investigation of the spectra of 2007sv, the similarity with the spectra of well-known SN impostors \citep[e.g. 1997bs and UGC2773-2009OT1,][]{2000PASP..112.1532V,2010AJ....139.1451S,2011ApJ...732...32F} is evident. 
However, some similarity can also be found with the spectra of genuine interacting SNe \citep[such as SNe~1999el and 1995G,][]{2002ApJ...573..144D,2002MNRAS.333...27P}. 
The comparisons shown in Section \ref{cfrsp} confirm that there are only subtle differences between spectra of LBV-like eruptions and genuine type IIn SNe, giving evidence that from the spectroscopic analysis alone it is sometimes tricky to discriminate between the two types of transients. 

However, from an in-depth inspection of our spectral sequence of 2007sv, we can obtain crucial information on this object. 
The increased total luminosity of H$_{\alpha}$ and the enhanced strength of the intermediate-velocity component in the late time spectra suggest that the material ejected in the outburst was interacting with the pre-existing CSM.
We also found that next to the expected narrow components, H$_\alpha$ and the Ca~II NIR triplet show broader wings (Figure \ref{cacomp}), suggesting an outflow of material at velocities comparable with those observed in SN ejecta. 
The maximum velocity registered for the outflowing material (dashed line in Figure \ref{cacomp}) is about 14000~km~s$^{-1}$ (although it is clear that the bulk of this material is expanding at much lower velocity, viz. $\sim$ 8000~km~s$^{-1}$). 
The detection of prominent broad spectral features from typical nucleosynthesis products observed in SN ejecta would be a more robust tool to distinguish between SNe and impostors. 
We do not detect any broad line of $\alpha$- or Fe-peak elements in the late-time spectra (approx. +80d) of 2007sv, except for the shallow absorption attributed to the Ca~II NIR triplet.
For this reason, the general spectral properties of 2007sv favor a non-terminal explosion scenario for this transient, although we have to admit that these can not be considered as conclusive proofs to unveil the nature of this interacting transient.

A more robust constraint can be derived from the photometric analysis. In some cases, impostors were unmasked through their erratic light curves. 
It is worth mentioning the optical transient observed during the period 2000-2009 in NGC~3432 \citep[aka 2000ch,][]{2004PASP..116..326W,2010MNRAS.408..181P} and also the 2009-2012 recurrent transient observed in NGC~7259 \citep[known as 2009ip,][]{2010AJ....139.1451S,2011ApJ...732...32F,2013ApJ...767....1P}. 
The latter was followed by a major eruption in mid-2012 that has been proposed to be a terminal SN explosion \citep[][and references therein]{2013MNRAS.tmp.2960S}. 
However, more frequently impostors reveals themselves through a single episode, characterized by a fast-evolving but regular light curve. 
A classical example is 1997bs \citep{2000PASP..112.1532V}, whose light curve shape is not trivially discernible from that of a regular SN. 
And the light curve of 2007sv is remarkably similar to that of 1997bs. 
However, the colours of 2007sv rapidly become redder as the object evolves, due to the decreasing temperature of the emitting region (this finding is confirmed by the temperature evolution inferred through black-body fits to the spectral continuum). 
The colour/temperature transition is observed to occur on much shorter timescales than in typical SNe IIn (see e.g. Fig. \ref{colcurves}). 

Finally, the peak luminosity still remains the most used method to discriminate between SNe and impostors. 
With a distance modulus of 31.38 $\pm$ 0.27 mag, we derive for 2007sv an absolute magnitude of M$_R = -14.25 \pm$ 0.38. 
This value is 4 to 6 magnitudes fainter than the absolute magnitudes typically measured in type IIn SNe \citep[e.g.][]{2002AJ....123..745R}.

\subsection{Which mechanisms can produce 2007sv-like events?}
The faint absolute magnitude at the discovery and the rapid colour (temperature) evolution provide robust evidence for the impostor nature for 2007sv, though some cases of faint transients exist in the literature that have been proposed to be true SNe. 
In fact, weak SNe can be produced via i) the core-collapse explosion of a peculiar SN through electron capture of a moderate-mass star; ii) the fall back event from a very massive star \cite[][and references therein]{2009ApJ...705L.138P}. 
Both mechanisms are believed to produce absolute light curves fainter than those observed in canonical SN types. 

In the former scenario the ONeMg stellar core of a 8-10 M$_{\odot}$ super-AGB star collapses generating a weak, low-energy event called electron-capture (EC) SN. 
The SN ejecta may eventually interact with an H-rich circumstellar environment generated by the stellar mass-loss during the super-AGB phase. 
As mentioned in Section \ref{cfrsp}, a promising candidate EC SN is SN~2008S \citep{2009MNRAS.398.1041B}, an object that shares some similarity with a SN impostor, but has a SN-like shaped light curve, a late-time light curve consistent with the $^{56}$Co decay. 
In the latter scenario, the collapse of a very massive star \citep[$>$ 25-30 M$_{\odot}$,][]{2003MNRAS.338..711Z} is followed by the fall-back of the inner stellar mantle onto the stellar core, generating eventually a black hole. 
In both scenarios a common feature is the faint absolute magnitude, which is generally due to the small amount of radioactive $^{56}$Ni in the ejecta. 
The presence of radioactive material in the ejecta can be revealed from the decline rate of the late-time SN light curve. 
However, in massive stars the interaction of the ejecta with the pre-existing CSM can induce a dramatic increase in the radiated energy, and cause significant deviations from the expected luminosity peak and light curve decline rate expected in the radioactive decays, making the detection of $^{56}$Co signatures problematic.

Another efficient mechanism proposed to explain transient events with a total radiated energy comparable with those of real SNe is the pulsational pair-instability in very massive stars. 
\citet{2007Natur.450..390W} showed that major instabilities produced by electron-positron pair production (pulsational-pair instability) cause the ejection of massive shells without necessarily unbinding the star (and hence without leading to a terminal SN explosion). 
These major mass-loss episodes might produce transients currently classified as SN impostors. In addition to this, when a new shell is ejected and collides with pre-existing material, the resulting radiated energy is comparable with that of a core-collapse SN (sometimes even one order of magnitude higher). 
If the impacting material is H-rich, the shell-shell collisions would produce a SN IIn-like spectrum and a slowly-evolving, luminous light curve that would make the transient practically indiscernible from true SNe IIn.  
All of this further complicates our attempts of discriminating SNe IIn from eruptive impostors.

A safe discrimination criterion would be the detection of the products of stellar and core-collapse explosive nucleosynthesis through the prominent $\alpha$-element lines in the nebular spectra. 
But in many SNe IIn the inner ejecta are covered by the H-rich interaction region sometimes for very long timescales (up to many years), making the detection of the $\alpha$-element spectral features difficult.

All the clues illustrated so far make us confident that 2007sv was not a terminal SN explosion, but very likely a major eruption mimicking the SN behaviour. 
If this is true, the progenitor star may have reached again a quiescent stage, returning to the pre-eruptive bolometric luminosity. 
This can be confirmed through an inspection of deep, high resolution images obtained years after the outburst, for example using the Hubble Space Telescope or the largest ground based telescopes which can deliver sub-arcsecond images.
The identification of the quiescent progenitor in such high quality images would be final evidence that the massive star producing 2007sv is still alive. 
Alternatively, a long timescale monitoring of 2007sv can eventually reveal further outbursts after the one registered in 2007, which would also prove that the 2007 episode was not the final stellar death. 
This strategy worked well for the 2000 transient observed in NGC~3432 \citep{2004PASP..116..326W} that was recovered after 8 years during a subsequent eruptive phase \citep{2010MNRAS.408..181P}. 

\subsection{Is 2007sv heralding a SN explosion?} 

\begin{figure}
\begin{center}
\includegraphics[scale=0.48]{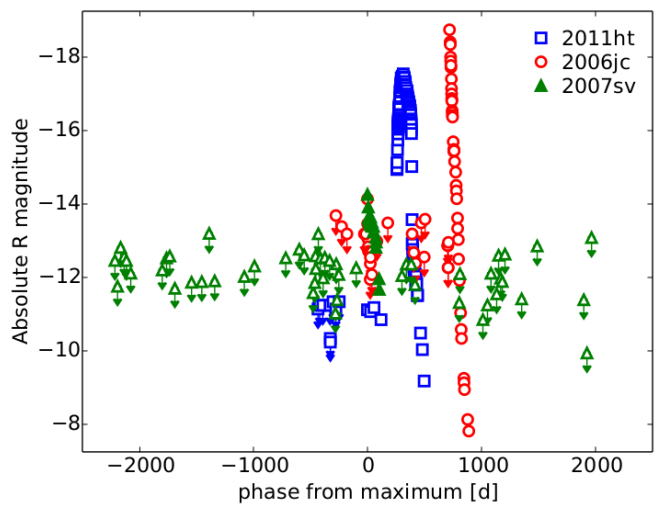}
\end{center}
\caption[Limits]{Plot of the photometric limits obtained from November 27, 2001 to January 07, 2014 and the absolute R-band light curve of 2007sv. The absolute light curve of the type Ibn SN~2006jc and the type IIn SN~2011ht are shown for comparison. The phases refer to the first recorded eruption.}
\label{limits}
\end{figure}

In order to identity possible further outbursts experienced by the progenitor of 2007sv, we analyzed a number of images of the transient site obtained before and after the 2007 event. 
These data were mostly collected by two of the co-authors of this paper (T.B. and G.D.), with a few additional observations performed with the 1.82~m Telescopio Copernico of the Asiago Observatory. 
These observations are listed in Table \ref{limtab} in Appendix \ref{lim}.
The unfiltered observations of T.B. and G.D. were scaled to R-band magnitudes using the magnitudes of references stars reported in Table \ref{localstandards}. 
These images of the 2007sv site span a period of over a decade. 
In this temporal window, we did not detect any further signature of a variable source at the position of 2007sv (see Figure \ref{limits}). 
The Pan-STARRS1 survey imaged this galaxy on 52 separate nights between Feb. 25, 2010 (MJD 55252.30) and May 17, 2013 (MJD 56429.26) in one of the filters $g_{\rm P1}r_{\rm P1}i_{\rm P1}z_{\rm P1}$ 
\citep[for a description of the Pan-STARRS1 3$\pi$ survey, see][]{2013ApJ...770..128I, 2013ApJS..205...20M}. No further detection of any outbursting activity was seen, and the magnitude limits of these individual epochs are typically 22.0, 21.6, 21.7, 21.4 and 19.3 respectively for  $g_{\rm P1}r_{\rm P1}i_{\rm P1}z_{\rm P1}$ \citep[as reported in][]{2013ApJ...770..128I}. These magnitudes are in the AB system as reported in \cite{2012ApJ...750...99T}.

Tracing the photometric history of a SN impostor has also another objective. 
As briefly mentioned in Section \ref{intro}, there is growing evidence that some interacting SNe may be preceded by large stellar eruptions (i.e. impostor events). 
A similar sequence of events has been proposed for a number of SNe, from the historical case of SN~2006jc discovered by K. Itagaki \citep{2006CBET..666....1N,2006CBET..666....2Y,2007Natur.447..829P} to a few recent type IIn SNe \citep[see][and references therein]{2013MNRAS.tmp.2960S}. 
In several cases, a lower luminosity outburst was observed few weeks before the brightest event (i.e. the putative SN). 
However, occasionally a larger time delay was observed between the two episodes (1-2 years). 
In Figure \ref{limits}, together with 2007sv, we also show two cases of SNe that were heralded by an outburst with a significant time delay. SN~2006jc is a stripped-envelope SN (of type Ibn) whose ejecta were seen to interact with He-rich CSM \citep{2007Natur.447..829P,2008MNRAS.389..113P,2007ApJ...657L.105F}. 
Its progenitor experienced an outburst 2 years before its explosion as a core-collapse SN, and the magnitude of this impostor (M$_R$ $\simeq$ -14) was comparable with that expected in an LBV eruption. 
SN~2011ht was initially classified as a SN impostor on the basis of its early spectral properties \citep{2011CBET.2851....2P}, and was later reclassified as a type IIn SN after a major spectral metamorphosis \citep{2011CBET.2903....1P}. 
However, the nature of SN~2011ht nature is not fully clarified. 
There are controversial interpretations on its SN-like observables \citep{2012ApJ...751...92R,2012ApJ...760...93H,2013MNRAS.431.2599M,2013ApJ...779L...8F}, since collisions among massive shells might still explain SN~2011ht without necessarily invoking a core-collapse. 
Interestingly, a posteriori, a weak transient has been observed about one year before the main episode \citep{2013ApJ...779L...8F}. 
This weak source, \citep[labelled as PSO J152.0441+51.8492 by][]{2013ApJ...779L...8F} was detected in archival data of the Panoramic Survey Telescope and Rapid Response System 1 (Pan-STARRS1) at an absolute magnitude M$_R \approx$ -11.8. \citet{2013ApJ...779L...8F} provided strong arguments that the two sources were physically related. 
Recent studies \citep[see][]{2014arXiv1401.5468O} claim that pre-SN eruptions are quite common. 
Nonetheless, so far only an handful of impostors with solid detections have been observed to be followed by what is believed to be a true SN explosion. 

The most intriguing issue is that, although there are some outliers with the absolute magnitudes brighter than -14 \citep{2014arXiv1401.5468O}, most pre-SN outbursts (those with robust detections) have absolute magnitudes close to or fainter than -14, nearly coincident with absolute magnitude of 2007sv. 
Although there was no further detection of a re-brightening of 2007sv before and after 2007, its overall photometric similarity with the precursors of SN~2006jc and other interacting SNe, may lead to the speculation that impostors such as 2007sv are instability episodes of massive stars (some of them in the LBV stage) that can be followed on short timescales (months to decades) by a terminal stellar explosion, as an ejecta-CSM interacting SN.

\section{Conclusions} \label{conclusions}
In this paper we have reported the results of our follow-up campaign of the transient 2007sv.
Our photometric monitoring spans a period of over 100d, whilst our spectroscopy data cover $\simeq$80d of the evolution of 2007sv. 
The spectra are largely dominated by a multi-component H$_\alpha$ line in emission. 
This spectral characteristic is common to both in SN impostors and in type IIn SNe. 

As we have illustrated, hough there are uncertainties on the observational constraints and the discrimination between true interacting SNe and SN impostors is often a tricky issue, the lack of broad lines from $\alpha$-elements, the fast colour/temperature evolutions, together with the relatively low velocity of the ejecta and the faint absolute magnitude at discovery (M$_R=-14.25$) support the SN impostor scenario for 2007sv, most likely a major eruption of an LBV. 
Although we do not have stringent pre-discovery limits, we claim that 2007sv was discovered soon after the outburst. 
Hence, we have adopted the discovery magnitude as an indicative guess for the peak magnitude.
This is supported by the rapid evolution of the colours and the temperature of the ejecta in the early phases supports this assumption.

However, some doubts still remain whether 2007sv was instead a very weak terminal explosion of a massive star. 
In absence of the detection of further outbursts or a future `real' SN explosion (as observed in other similar transients), the most promising method to definitely rule out the possibility that 2007sv was a faint interacting core-collapse SN is by obtaining deep and high spatial resolution images of the transient's site (e.g. with HST), in the attempt to detect some signatures from the surviving star. 
This method, that has been successfully tested to find the progenitors of a number of CC-SNe \citep[see][and references therein]{2009ARA&A..47...63S} and may well give the final answer to the enigma of 2007sv.

\section*{Acknowledgements} \label{ackn}
Based on observations made with:
- The Cima Ekar 1.82~m telescope of the INAF-Astronomical Observatory of Padua, Italy; \\
- The Liverpool Telescope operated on the island of La Palma by Liverpool John Moores University in the Spanish Observatorio del Roque de los Muchachos of the Instituto de Astrofisica de Canarias with financial support from the UK Science and Technology Facilities Council. \\
- the Nordic Optical Telescope (NOT), operated by the Nordic Optical Telescope Scientific Association at the Observatorio del Roque de los Muchachos, La Palma, Spain, of the Instituto de Astrofisica de Canarias. \\
- The William Herschel Telescope (WHT) operated on the island of La Palma by the Isaac Newton Group in the Spanish Observatorio del Roque de los Muchachos of the Instituto de Astrofsica de Canarias. \\
- The Italian Telescopio Nazionale Galileo (TNG) operated on the island of La Palma by the Fundacion Galileo Galilei of the INAF (Istituto Nazionale di Astrofisica) at the Spanish Observatorio del Roque de los Muchachos of the Instituto de Astrofisica de Canarias. \\
- The Hobby-Eberly Telescope (HET), joint project of the University of Texas at Austin, Pennsylvania State University, Stanford University, Ludwig-Maximilians-Universit{\"a}t M{\"u}nchen, and Georg-August-Universit{\"a}t G{\"o}ttingen.  \\
- The Calar Alto 2.2~m telescope operated at the Centro Astronomico Hispano Aleman (CAHA) at Calar Alto, owned and operated jointly by the Max-Planck-Institut f{\"u}r Astronomie in Heidelberg, Germany, and the Instituto de Astrofisica de Andalucia in Granada, Spain \\
We thank P.~Corelli (Mandi observatory, Pagnacco, UD, Italy) for the observations of (SN) 2007sv. \\
Based also on observations performed at the Nordic Optical Telescope (Proposal number 49-016, PI: F. Taddia), La Palma, Spain. \\
This work is part of the European supernova collaboration involved in the ESO-NTT large programme 184.D-1140 led by Stefano Benetti. \\
L.T., A.P., S.B., E.C., A.H. and M.T. are partially supported by the PRIN-INAF 2011 with the project `Transient Universe: from ESO Large to PESSTO`. \\
N.E.R. is supported by the MICINN grant AYA2011-24704/ESP, by the ESF EUROCORES Program EuroGENESIS (MICINN grant EUI2009-04170), by SGR grants of the Generalitat de Catalunya, and by EU-FEDER funds. \\
N.E.R. acknowledges the support from the European Union Seventh Framework Programme (FP7/2007-2013) under grant agreement n. 267251. \\
S.T. acknoledges support by TRR 33 `The Dark Universe' of the German Reasearch Foundation (DFG). \\
The research of JRM is supported through a Royal Society Research Fellowship. \\
The research leading to these results has received funding from the European Research Council under the European Union's Seventh Framework Programme (FP7/2007-2013)/ERC Grant agreement n$^{\rm o}$ [291222] (PI: S.~J.~Smartt). \\
FB acknowledges support from FONDECYT through Postdoctoral grant 3120227 and from Project IC120009 `Millennium Institute of Astrophysics (MAS)' of the Iniciativa Cient�fica Milenio del Ministerio de Econom�a, Fomento y Turismo de Chile. \\
The Pan-STARRS1 Surveys (PS1) have been made possible through contributions of the Institute for Astronomy, the University of Hawaii, the Pan-STARRS Project Office, the Max-Planck Society and its participating institutes, the Max Planck Institute for Astronomy, Heidelberg and the Max Planck Institute for Extraterrestrial Physics, Garching, The Johns Hopkins University, Durham University, the University of Edinburgh, Queen's University Belfast, the Harvard-Smithsonian Center for Astrophysics, the Las Cumbres Observatory Global Telescope Network Incorporated, the National Central University of Taiwan, the Space Telescope Science Institute, the National Aeronautics and Space Administration under Grant No. NNX08AR22G issued through the Planetary Science Division of the NASA Science Mission Directorate, the National Science Foundation under Grant No. AST-1238877, the University of Maryland, and Eotvos Lorand University (ELTE).

\appendix
\section{Photometric limits of 2007sv} \label{lim}

\begin{table*}
\begin{minipage}{175mm}
 \caption[Limits of 2007sv]{Optical detection limits for 2007sv. No source was observed in the position of the transient at any epochs reported in the Table.}
 \label{limtab}
 \begin{tabular}{@{}cccc@{}}
 \hline
    Date  &    MJD   & R         & Instrument \\
 \hline
 20011127 & 52240.11 & $>$ 18.96 & Apogee AP7 \\ 
 20011220 & 52263.09 & $>$ 19.66 & Apogee AP7 \\
 20020116 & 52290.20 & $>$ 18.60 & Apogee AP7 \\ 
 20020215 & 52320.81 & $>$ 18.85 & Apogee AP7 \\ 
 20020307 & 52340.86 & $>$ 18.96 & Apogee AP7 \\ 
 20020414 & 52378.95 & $>$ 19.30 & Apogee AP7 \\ 
 20030115 & 52654.91 & $>$ 19.22 & Apogee AP7 \\ 
 20030222 & 52692.91 & $>$ 18.89 & Apogee AP7 \\ 
 20030323 & 52721.03 & $>$ 18.84 & Apogee AP7 \\ 
 20030506 & 52765.91 & $>$ 19.72 & Apogee AP7 \\ 
 20030930 & 52912.17 & $>$ 19.55 & Apogee AP7 \\ 
 20031229 & 53002.03 & $>$ 19.54 & Apogee AP7 \\ 
 20040302 & 53066.83 & $>$ 18.22 & Apogee AP7 \\ 
 20040419 & 53114.94 & $>$ 19.51 & Apogee AP7 \\ 
 20050105 & 53375.00 & $>$ 19.39 & Apogee AP7 \\ 
 20050403 & 53463.98 & $>$ 19.11 & Apogee AP7 \\ 
 20060103 & 53738.05 & $>$ 18.89 & SX         \\
 20060305 & 53899.87 & $>$ 18.81 & Apogee AP7 \\ 
 20060501 & 53856.89 & $>$ 18.65 & Apogee AP7 \\ 
 20060831 & 53978.90 & $>$ 19.84 & SX         \\
 20060915 & 53993.00 & $>$ 19.56 & SX         \\
 20060920 & 53998.95 & $>$ 18.83 & SX         \\
 20060923 & 54001.87 & $>$ 19.18 & SX         \\
 20061015 & 54023.78 & $>$ 18.23 & SX         \\
 20061103 & 54042.08 & $>$ 18.89 & SX         \\
 20061126 & 54065.91 & $>$ 19.43 & SX         \\
 20061223 & 54081.06 & $>$ 18.88 & SX         \\
 20070122 & 54122.88 & $>$ 18.96 & SX         \\
 20070314 & 54173.86 & $>$ 20.39 & SX         \\
 20070321 & 54180.88 & $>$ 19.05 & Apogee AP7 \\ 
 20070328 & 54187.84 & $>$ 20.03 & SX         \\
 20070413 & 54203.85 & $>$ 19.34 & SX         \\
 20070913 & 54356.09 & $>$ 19.16 & SX         \\
 20081017 & 54756.97 & $>$ 19.37 & SX         \\
 20081125 & 54795.98 & $>$ 19.05 & SX         \\
 20090107 & 54838.84 & $>$ 19.06 & SX         \\
 20090209 & 54871.83 & $>$ 19.62 & SX         \\
 20100304 & 55259.84 & $>$ 20.10 & Apogee AP7 \\ 
 20100310 & 55265.80 & $>$ 19.32 & SX         \\
 20100928 & 55467.13 & $>$ 20.57 & SX         \\
 20101107 & 55507.13 & $>$ 20.17 & SX         \\
 20101207 & 55537.95 & $>$ 19.31 & SX         \\
 20110126 & 55587.84 & $>$ 19.87 & SX         \\
 20110212 & 55604.81 & $>$ 18.82 & Apogee AP7 \\
 20110302 & 55622.86 & $>$ 19.44 & SX         \\
 20110316 & 55636.87 & $>$ 19.53 & SX         \\ 
 20110409 & 55660.94 & $>$ 18.79 & Apogee AP7 \\
 20110903 & 55807.07 & $>$ 20.01 & SX         \\ 
 20120117 & 55943.21 & $>$ 18.56 & Apogee AP7 \\
 20130228 & 56351.89 & $>$ 20.03 & ARTEMIS    \\
 20130315 & 56379.86 & $>$ 21.48 & ARTEMIS    \\
 20130320 & 56371.10 & $>$ 21.88 & AFOSC      \\ 
 20130507 & 56419.92 & $>$ 18.34 & Apogee AP7 \\
 20131205 & 56631.19 & $>$ 22.15 & AFOSC      \\
 20140107 & 56664.99 & $>$ 21.29 & AFOSC      \\
 \hline
 \end{tabular}

 \medskip
 The observations provided by T.~B. (with a C-14 Celestron Schmidt Cassegrain reflector and an Apogee AP7 CCD camera at the Coddenham Astronomical Observatory, Suffolk, United Kingdom) and G.~D. (with a 0.32~m f/3.1 reflector and a Starlight Xpress MX716 CCD camera at Moonbase Observatory (Akersberga, Sweden)) were unfiltered images, whith magnitudes rescaled to R-band. \\ 
 Multi-band observations were obtained on March 20, 2013, with the following additional detection limits: U $>$ 19.93, B $>$ 20.94, V $>$ 21.02, I $>$ 21.26.
\end{minipage}
\end{table*}

\label{lastpage}


\begin{thebibliography}{99}
\bibitem[Blanc et al.(2005)]{2005ATel..630....1B} Blanc, N., Bongard, S., Copin, Y., et al.\ 2005, The Astronomer's Telegram, 630, 1 
\bibitem[Berger et al.(2009)]{2009ApJ...699.1850B}  Berger, E., Soderberg, A. M., Chevalier, R. A.\ 2009, ApJ, 699, 1850
\bibitem[Bond et al.(2009)]{2009ApJ...695L.154B} Bond, H. E., Bedin, L. R., Bonanos, A. Z., Humphreys, R. M,; Monard, L. A. G. B., Prieto, J. L.; Walter, F. M.\ 2009, ApJ, 695, L154
\bibitem[Botticella et al.(2009)]{2009MNRAS.398.1041B} Botticella, M.~T., Pastorello, A., Smartt, S.~J., et al.\ 2009, MNRAS, 398, 1041
\bibitem[Chevalier \& Fransson(1994)]{1994ApJ...420..268C} Chevalier, R. A., Fransson, C. 1994, ApJ, 420, 268
\bibitem[Corwin et al.(1985)]{1985sgcc.book.....C} Corwin, H.~G., de Vaucouleurs, A., \& de Vaucouleurs, G.\ 1985, University of Texas Monographs in Astronomy, Austin: University of Texas, 1985
\bibitem[Dessart et al.(2009)]{2009MNRAS.394...21D} Dessart, L., Hillier, D.~J., Gezari, S., Basa, S., \& Matheson, T.\ 2009, MNRAS, 394, 21 
\bibitem[Di Carlo et al.(2002)]{2002ApJ...573..144D} Di Carlo, E., Massi, F., Valentini, G., et al.\ 2002, ApJ, 573, 144 
\bibitem[Duszanowicz et al.(2007)]{2007CBET.1182....1D} Duszanowicz, G., Boles, T., \& Corelli, P.\ 2007, Central Bureau Electronic Telegrams, 1182, 1
\bibitem[Foley et al.(2007)]{2007ApJ...657L.105F} Foley, R. J., Smith, N., Ganeshalingam, M., Li, W., Chornock, R., Filippenko, A. V. 2007, ApJ, 657, L105
\bibitem[Foley et al.(2011)]{2011ApJ...732...32F} Foley, R. J., Berger, E., Fox, O., Levesque, E. M., Challis, P. J., Ivans, I. I., Rhoads, J. E., Soderberg, A. M.\ 2011, ApJ, 732, 32
\bibitem[Fraser et al.(2013)]{2013ApJ...779L...8F} Fraser, M., Magee, M., Kotak, R., et al.\ 2013, ApJL, 779, L8 
\bibitem[Gal-Yam \& Leonard(2009)] {2009Natur.458..865G} Gal-Yam, A., Leonard, D. C.\ 2009, Nature, 458, 865
\bibitem[Grebel(2001)]{2001ApSSS.277..231G} Grebel, E.~K.\ 2001, Astrophysics and Space Science Supplement, 277, 231
\bibitem[Habergham et al.(2014)]{2014MNRAS.441.2230H} Habergham, S.~M., Anderson, J.~P., James, P.~A., \& Lyman, J.~D.\ 2014, MNRAS, 441, 2230
\bibitem[Huchtmeier \& Skillman(1998)]{1998A&AS..127..269H} Huchtmeier, W.~K., \& Skillman, E.~D.\ 1998, A\&Aps, 127, 269
\bibitem[Humphreys \& Davidson(1994)]{1994PASP..106.1025H} Humphreys, R.~M., \& Davidson, K.\ 1994, PASP, 106, 1025
\bibitem[Humphreys et al.(2012)]{2012ApJ...760...93H} Humphreys, R.~M., Davidson, K., Jones, T. J., Pogge, R. W., Grammer, S. H., Prieto, J. L., Pritchard, T. A. 2012, ApJ, 760, 93
\bibitem[Inserra et al.(2013)]{2013ApJ...770..128I} Inserra, C., Smartt, S.~J., Jerkstrand, A., et al.\ 2013, ApJ, 770, 128 
\bibitem[Khan et al.(2010)]{2010ApJ...715.1094K} Khan, R., Stanek, K. Z., Prieto, J. L., Kochanek, C. S., Thompson, T. A., \& Beacom, J. F.\ 2010, ApJ, 715, 1094
\bibitem[Kochanek(2011)]{2011ApJ...741...37K} Kochanek, C. S.\ 2011, ApJ, 741, 37
\bibitem[Kotak \& Vink(2006)]{kot06} Kotak, R., Vink, J.\ 2006, A\&A, 460, L5
\bibitem[Itagaki et al.(2006)]{2006IAUC.8762....1I} Itagaki, K., Nakano, S., Puckett, T., et al.\ 2006, IAU Circ., 8762, 1
\bibitem[Landolt(1992)]{1992AJ....104..340L} Landolt, A.~U.\ 1992, AJ, 104, 340
\bibitem[Magnier et al.(2013)]{2013ApJS..205...20M} Magnier, E.~A., Schlafly, E., Finkbeiner, D., et al.\ 2013, ApJs, 205, 20 
\bibitem[Margutti et al.(2014)]{2013arXiv1306.0038M} Margutti, R., et al. 2014, ApJ, 780, 21
\bibitem[Mattila et al.(2008)]{2008MNRAS.389..141M} Mattila, S., Meikle, W. P. S., Lundqvist, P. et al. 2008, MNRAS,  389, 141
\bibitem[Maund et al.(2006)]{2006MNRAS.369..390M} Maund, J.~R., Smartt, S.~J., Kudritzki, R.-P., et al.\ 2006, MNRAS, 369, 390 
\bibitem[Mauerhan et al.(2013)]{2013MNRAS.431.2599M} Mauerhan, J.~C., Smith, N., Silverman, J.~M., et al.\ 2013, MNRAS, 431, 2599
\bibitem[Nakano et al.(2006)]{2006CBET..666....1N} Nakano, S., Itagaki, K., Puckett, T., Gorelli, R.\ 2006, CBET, 666, 1
\bibitem[Nozawa et al.(2008)]{2008ApJ...684.1343N} Nozawa, T., Kozasa, T., Tominaga, N. et al., 2008, ApJ, 684, 1343 
\bibitem[Ofek et al.(2013)]{2013Natur.494...65O} Ofek, E. O. et al.\ 2013, Nature, 494, 65
\bibitem[Ofek et al.(2014)]{2014arXiv1401.5468O} Ofek et al. 2014, ApJ submitted ()arXiv:1401.5468
\bibitem[Pastorello et al.(2002)]{2002MNRAS.333...27P} Pastorello, A., Turatto, M., Benetti, S. et al. 2002, MNRAS, 333, 27
\bibitem[Pastorello et al.(2007)]{2007Natur.447..829P} Pastorello, A., Smartt, S.~J., Mattila, S., et al.\ 2007, Nature, 447, 829
\bibitem[Pastorello et al.(2008)]{2008MNRAS.389..113P} Pastorello, A., Mattila, S., Zampieri, L., et al.\ 2008, MNRAS, 389, 113
\bibitem[Pastorello et al.(2010)]{2010MNRAS.408..181P} Pastorello, A., Botticella, M. T., Trundle, C., et al. 2012, MNRAS, 408, 181
\bibitem[Pastorello et al.(2011)]{2011CBET.2851....2P} Pastorello, A., Stanishev, V., Smartt, S. J., Fraser, M.; Lindborg, M.\ 2011, CBET, 2851, 2
\bibitem[Pastorello et al.(2013)]{2013ApJ...767....1P} Pastorello, A., Cappellaro, E., Inserra, C., et al.\ 2013, ApJ, 767, 1 
\bibitem[Pettini \& Pagel(2004)]{2004MNRAS.348L..59P} Pettini, M., \& Pagel, B.~E.~J.\ 2004, MNRAS, 348, L59 
\bibitem[Pilyugin et al.(2004)]{2004A&A...425..849P} Pilyugin, L.~S., V{\'{\i}}lchez, J.~M., \& Contini, T.\ 2004, A\&A, 425, 849
\bibitem[Prieto et al.(2008)]{2008ApJ...681L...9P} Prieto, J. L., Kistler, M. D., Thompson, T. A. et al. 2008, ApJ, 681, L9
\bibitem[Prieto et al.(2011)]{2011CBET.2903....1P} Prieto, J.~L., McMillan, R., Bakos, G., \& Grennan, D.\ 2011, Central Bureau Electronic Telegrams, 2903, 1
\bibitem[Pumo et al.(2009)]{2009ApJ...705L.138P} Pumo, M.~L., Turatto, M., Botticella, M.~T., et al.\ 2009, ApJl, 705, L138 
\bibitem[Richardson et al.(2002)]{2002AJ....123..745R} Richardson, D.,; Branch, D., Casebeer, D., Millard, J., Thomas, R. C., Baron, E. 2002, AJ, 123, 745
\bibitem[Roming et al.(2012)]{2012ApJ...751...92R} Roming, P. W. A., Pritchard, T. A., Prieto, J. L.\ 2012, ApJ, 751, 92
\bibitem[Schlafly \& Finkbeiner(2011)]{2011ApJ...737..103S} Schlafly, E.~F., \& Finkbeiner, D.~P.\ 2011, ApJ, 737, 103
\bibitem[Schlegel et al.(1998)]{1998ApJ...500..525S} Schlegel, D.~J., Finkbeiner, D.~P., \& Davis, M.\ 1998, ApJ, 500, 525
\bibitem[Smartt(2009)]{2009ARA&A..47...63S} Smartt, S. J. 2009, ARA$\&$A, 47, 63
\bibitem[Szczygiel et al.(2012)]{2012ApJ...750...77S} Szczygiel, D. M., Prieto, J. L., Kochanek, C. S., Stanek, K. Z., Thompson, T. A., Beacom, J. F., Garnavich, P. M., Woodward, C. E.\ 2012, ApJ, 750, 77 
\bibitem[Smith (2008)]{2008Natur.455..201S} Smith, N. 2008, Nature, 455, 201
\bibitem[Smith et al.(2008)]{2008ApJ...680..568S} Smith, N., Foley, R. J., Filippenko, A. V. 2008, ApJ, 687, 1208
\bibitem[Smith et al.(2009)]{2009ApJ...697L..49S} Smith, N., Ganeshalingam, M., Chornock, R. et al. 2009, ApJ, 679, L49
\bibitem[Smith et al.(2010)]{2010AJ....139.1451S} Smith, N., Miller, A., Li, W., et al. 2010, AJ, 139, 1451
\bibitem[Smith et al.(2011)]{2011MNRAS.415..773S} Smith, N., Li, W., Silverman, J. M., Ganeshalingam, M., Filippenko, A. V. 2011, MNRAS, 415, 773
\bibitem[Smith et al.(2013)]{2013MNRAS.434.2721S} Smith, N., Mauerhan, J.~C., Kasliwal, M.~M., \& Burgasser, A.~J.\ 2013, MNRAS, 434, 2721 
\bibitem[Smith et al.(2014)]{2013MNRAS.tmp.2960S} Smith, N., Mauerhan, J.~C., Prieto, J.\ 2014, MNRAS in press (arXiv:1308.0112) 
\bibitem[Stritzinger et al.(2012)]{2012ApJ...756..173S} Stritzinger, M., Taddia, F., Fransson, C., et al.\ 2012, ApJ, 756, 173
\bibitem[Taddia et al.(2013)]{2013A&A...558A.143T} Taddia, F., Sollerman, J., Razza, A., et al.\ 2013, A\&A, 558, A143 
\bibitem[Tammann(1994)]{1994ESOC...49....3T} Tammann, G.~A.\ 1994, European Southern Observatory Conference and Workshop Proceedings, 49, 3
\bibitem[Thompson et al.(2009)]{2009ApJ...705.1364T} Thompson, T.~A., Prieto, J.~L., Stanek, K.~Z., et al.\ 2009, ApJ, 705, 1364 
\bibitem[Tominaga et al.(2013)]{2013ApJ...771L..12T} Tominaga, N., Blinnikov, S. I., Nomoto, K. 2013, ApJ, 771, L12
\bibitem[Tonry et al.(2012)]{2012ApJ...750...99T} Tonry, J.~L., Stubbs, C.~W., Lykke, K.~R., et al.\ 2012, ApJ, 750, 99 
\bibitem[Turatto et al.(1993)]{1993MNRAS.262..128T} Turatto, M., Cappellaro, E., Danziger, I.~J., et al.\ 1993, MNRAS, 262, 128 
\bibitem[van den Bergh(1960)]{1960ApJ...131..215V} van den Bergh, S.\ 1960, ApJ, 131, 215
\bibitem[Van Dyk et al.(2000)]{2000PASP..112.1532V} Van Dyk, S.~D., Peng, C.~Y., King, J.~Y., et al.\ 2000, PASP, 112, 1532 
\bibitem[Van Dyk et al.(2012)]{2012ASSL..384..249V} Van Dyk, S. D., Li, W. 2012, {\it Eta Carinae and the Supernova Impostors}, Astrophysics and Space Science Library, Volume 384. ISBN 978-1-4614-2274-7. Springer Science+Business Media, LLC, p. 249
\bibitem[Wagner et al.(2004)]{2004PASP..116..326W} Wagner, R.~M., Vrba, F.~J., Henden, A.~A., et al.\ 2004, PASP, 116, 326
\bibitem[Wanajo et al.(2009)]{2009ApJ...695..208W} Wanajo, S., Nomoto, K., Janka, H.-T., Kitaura, F.~S., M\"uller, B.\ 2009, ApJ, 695, 208
\bibitem[Wesson et al.(2010)]{2010MNRAS.403..474W} Wesson, R., Barlow, M. J., Ercolano, B. et al. 2010, MNRAS, 403, 474
\bibitem[Wolf(1989)]{1989A&A...217...87W} Wolf, B.\ 1989, A\&A, 217, 87
\bibitem[Woosley et al.(2007)]{2007Natur.450..390W} Woosley, S.~E., Blinnikov, S., \& Heger, A.\ 2007, Nature, 450, 390
\bibitem[Yamaoka et al.(2006)]{2006CBET..666....2Y}Yamaoka, H., Nakano, S., Itagaki, K.\ 2006, CBET, 666, 2
\bibitem[Zampieri et al.(2003)]{2003MNRAS.338..711Z} Zampieri, L., Pastorello, A., Turatto, M., et al.\ 2003, MNRAS, 338, 711 
\end{thebibliography}
\end{document}